\documentclass[prd,twocolumn,amssymb,amsmath]{revtex4-1}
\pdfoutput=1
\usepackage{CJK}
\usepackage{graphicx}
\usepackage{amsmath,amssymb}
\usepackage{amsthm}
\usepackage{enumerate}
\usepackage{txfonts}
\usepackage{hyperref}
\usepackage{color,amsxtra,amsfonts,bm}

\begin{document}

\title{Gauge redundancy-free formulation of compact QED with dynamical matter for quantum and classical computations}

\date{\today}

\author{Julian Bender}
\address{Max-Planck-Institut f\"ur Quantenoptik, Hans-Kopfermann-Stra\ss e 1, 85748 Garching, Germany,}
\address{Munich Center for Quantum Science and Technology (MCQST), Schellingstr. 4, D-80799 M\"unchen, Germany,}

\author{Erez Zohar}
\address{Racah Institute of Physics, The Hebrew University of Jerusalem, Jerusalem 91904, Givat Ram, Israel.}

\begin{abstract}
We introduce a way to express compact quantum electrodynamics with dynamical matter on two- and three-dimensional spatial lattices in a gauge redundancy-free manner while preserving translational invariance. By transforming to a rotating frame, where the matter is decoupled from the gauge constraints, we can express the gauge field operators in terms of dual operators. In two space dimensions, the dual representation is completely free of any local constraints. In three space dimensions, local constraints among the dual operators remain but involve only the gauge field degrees of freedom (and not the matter degrees of freedom). These formulations, which reduce the required Hilbert space dimension, could be useful for both numerical (classical) Hamiltonian computations and quantum simulation or computation.
\end{abstract}

\maketitle

\section{Introduction}

Gauge theories appear in several important contexts in modern physics; most fundamentally, as the mechanism mediating the interactions in the standard model of particle physics \cite{peskin_introduction_1995,langacker_standard_2011}. The role they play in physics could not be underestimated, but the same could be said about the challenge they impose due to their strongly coupled nature that requires the use of non-perturbative techniques. For decades, lattice gauge theories \cite{wilson_confinement_1974,kogut_hamiltonian_1975} have provided a very fruitful toolbox for the study of gauge theories, by discretizing them on a lattice, either as a regularization scheme or as a computational approach, with quantum Monte Carlo. However, those numerical computations, performed in a Wick-rotated, Euclidean spacetime, still impose two major restrictions en route to a full understanding of gauge theories; one, is the impossibility to directly observe real-time dynamics in Euclidean spacetime, and the other is the well-known sign problem \cite{troyer_computational_2005}, which blocks the way to the study of several important physical phases in gauge theories, for example in quantum chromodynamics with a finite chemical potential \cite{mclerran_physics_1986,fukushima_phase_2011}.

Lattice gauge theories can either be formulated on a discretized spacetime \cite{wilson_confinement_1974}, or discretized space \cite{kogut_hamiltonian_1975}. The first approach is useful for the path integral, action formalism, which has been used widely for successful Monte-Carlo computations \cite{aoki2020flag}. The latter is well suited for a Hamiltonian approach. Besides the formulation proposed by Kogut and Susskind \cite{kogut_hamiltonian_1975} (and truncations thereof \cite{zohar_formulation_2015}), there are other formulations such as the quantum link model \cite{horn1981,orland_lattice_1990,brower_qcd_1999} and the prepotential formalism \cite{mathur_harmonic_2005}. 

In the last few years, these Hamiltonian formulations have attracted more attention due to the development of new techniques in quantum many-body physics which might help overcome the aforementioned problems in the action formalism. One method involves quantum simulation \cite{feynman_simulating_1982, Cirac_goals_2012} of lattice gauge theories, that is, their mapping into quantum devices that can be controlled and manipulated as laboratory table-top experiments - cold atoms, trapped ions, superconducting qubits or other atomic, optical or solid-state devices \cite{wiese_ultracold_2013,zohar_quantum_2016,dalmonte_lattice_2016,banuls_simulating_2019,gorg_realization_2019,schweizer_floquet_2019,mil_realizing_2019,yang_observation_2020}. Another approach uses classical computation with variational ansatz states (in particular tensor networks) \cite{dalmonte_lattice_2016,banuls_simulating_2019,banuls_review_2019}. The idea is to find classes of states which are efficiently computable but at the same time capture the relevant features of the theory under consideration. 

Physical states in gauge theories need to satisfy local constraints (Gauss' laws). This has implications both for quantum simulations and for classical computations with variational states. For the former, one has to make sure that experimental errors do not lead to a violation of gauge invariance \cite{halimeh_robustness_2020,halimeh_gauge-symmetry_2020}. For the latter, the local constraints need to be incorporated in the variational ansatz. While in some cases these constraints can be useful for the construction of variational states (e.g. to build tensor network ansatz states), it makes it in general more difficult to find suitable ansatz states.

An alternative approach is to find a formulation of lattice gauge theories directly in terms of gauge-invariant variables. In a quantum simulation thereof, gauge invariance would be robust against experimental errors. For the construction of variational states, one could choose from a wider class of ansatz states due to the absence of constraints (e.g. generalizations of Gaussian states \cite{sala_variational_2018,PhysRevResearch.2.043145}). Also since the required resources are reduced by going from the full Hilbert space to the physical subspace, it seems worthwhile to find gauge-invariant formulations while preserving as many symmetries of the original formulation as possible. So far, works in this direction \cite{drell_quantum_1979,cobanera2011bondalgebraic,kaplan_gauss_2018,unmuthyockey2019gaugeinvariant,haase_resource_2020} have mainly focused on ($1+1$)-dimensional lattice gauge theories with dynamical matter and pure gauge theories in $2+1$ dimensions. The local gauge constraints have also been used to eliminate the matter degrees of freedom from the theory, including the Abelian Higgs model, where unitary gauge fixing is used \cite{meurice2020abeliangauge,fradkin1979higgs}, as well as a recent extension to fermionic scenarios \cite{zohar2018eliminating,zohar2020removing}.  

In this work, we discuss compact quantum electrodynamics (cQED) in two and three space dimensions, including dynamical gauge fields and matter (either fermionic or bosonic) and show how one can express it in terms of dual variables that reduce the number of local constraints: in two space dimensions, no local constraints are left, while in three dimensions, those constraints do not involve the matter. The formulation preserves translational invariance and is based on the decomposition of lattice vector fields into longitudinal and transverse parts, allowing us to decouple the matter from the constraints as a first step.

In the next section, we introduce the model in two space dimensions and review some lattice vector calculus basics that are required for our procedure. In section \ref{decoupsec}, we proceed to introduce a unitary transformation that allows one to eliminate the matter from the constraints or, in other words, split the longitudinal and transverse components of the electric field. In section \ref{dualformulation}, we show how to formulate the transformed model in terms of dual variables, as in other dual schemes, which results in non-local gauge-matter interactions; we proceed and introduce another, new set of dual variables that makes this interaction local again and leads to Coulomb-type interactions for both the gauge field and the matter degrees of freedom. Finally, in section \ref{threedimensions}, we discuss the generalization to three space dimensions and differences to the two dimensional case. 

Throughout the paper, we sum over doubly repeated indices, unless specified otherwise.

\section{Review of the Relevant Models} \label{review}

First, we shall restrict our discussion to a two dimensional ($2+1$d) system, in which a compact $U(1)$ gauge field \cite{kogut_introduction_1979} is coupled to some dynamical matter degrees of freedom: either fermions, as in Kogut and Susskind's formulation \cite{kogut_hamiltonian_1975} or other forms of matter, as described below. The following discussion is valid for both periodic and open boundary conditions. We will use periodic boundary conditions (a torus); the differences, which arise in the case of open boundary conditions, are mentioned throughout the paper. For periodic boundary conditions we assume a square lattice of extent $N \times N$, for open boundary conditions we consider $(N+1) \times (N+1)$ lattice sites, i.e. a lattice made of $N \times N$ plaquettes.

\subsection{The Lattice}\label{latsec}

The matter degrees of freedom reside on the sites of the lattice, labeled for periodic boundary conditions by integers $\mathbf{x}=\left(x_1,x_2\right)\in \{0,..,N-1 \}^2$ (for open boundary conditions $\left(x_1,x_2\right)\in \{0,..,N \}^2$), while the gauge fields - on the links, labeled by the site $\mathbf{x}$ from which they emanate and a direction $i=1,2$ to which they extend. The link labeled by $\mathbf{x},i$ connects the site $\mathbf{x}$ with the site $\mathbf{x}+\hat{\mathbf{e}}_i$, where $\hat{\mathbf{e}}_i$ is a unit vector pointing in the positive $i$ direction.

We consider three different kinds of lattice fields: Fields $f\left(\mathbf{x}\right)$, residing on the lattice sites $\mathbf{x}$ (such as matter fields or scalar fields), vector fields $\mathbf{F}\left(\mathbf{x}\right)$, 
whose components $F_{i}\left(\mathbf{x}\right)$ reside on the links of the lattice (such as vector potentials and electric fields) and pseudovector fields $B(\mathbf{x})$ (such as the magnetic field), residing on the plaquettes (denoted by the site $\mathbf{x}$ at the bottom left corner).

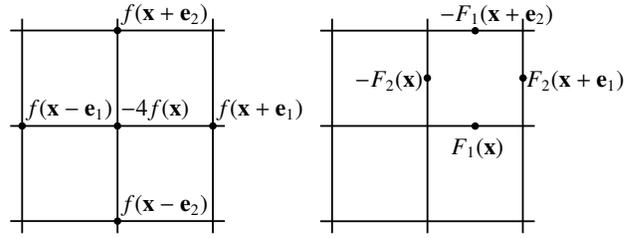
\begin{figure} 
	\centering
	\def\svgwidth{\columnwidth}
	\begingroup%
	\makeatletter%
	\providecommand\color[2][]{%
		\errmessage{(Inkscape) Color is used for the text in Inkscape, but the package 'color.sty' is not loaded}%
		\renewcommand\color[2][]{}%
	}%
	\providecommand\transparent[1]{%
		\errmessage{(Inkscape) Transparency is used (non-zero) for the text in Inkscape, but the package 'transparent.sty' is not loaded}%
		\renewcommand\transparent[1]{}%
	}%
	\providecommand\rotatebox[2]{#2}%
	\newcommand*\fsize{\dimexpr\f@size pt\relax}%
	\newcommand*\lineheight[1]{\fontsize{\fsize}{#1\fsize}\selectfont}%
	\ifx\svgwidth\undefined%
	\setlength{\unitlength}{204.57755568bp}%
	\ifx\svgscale\undefined%
	\relax%
	\else%
	\setlength{\unitlength}{\unitlength * \real{\svgscale}}%
	\fi%
	\else%
	\setlength{\unitlength}{\svgwidth}%
	\fi%
	\global\let\svgwidth\undefined%
	\global\let\svgscale\undefined%
	\makeatother%
	\begin{picture}(1,0.36483935)%
	\lineheight{1}%
	\setlength\tabcolsep{0pt}%
	\put(0,0){\includegraphics[width=\unitlength,page=1]{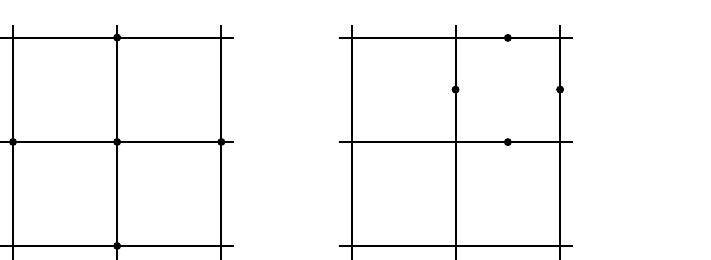}}%
	\put(0.17006867,0.18034996){\makebox(0,0)[lt]{\lineheight{1.25}\smash{\begin{tabular}[t]{l}$-4f(\mathbf{x})$\end{tabular}}}}%
	\put(0.31688353,0.18034992){\makebox(0,0)[lt]{\lineheight{1.25}\smash{\begin{tabular}[t]{l}$f(\mathbf{x}+\mathbf{e}_1)$\end{tabular}}}}%
	\put(0.17030075,0.32722379){\makebox(0,0)[lt]{\lineheight{1.25}\smash{\begin{tabular}[t]{l}$f(\mathbf{x}+\mathbf{e}_2)$\end{tabular}}}}%
	\put(0.17026817,0.03393643){\makebox(0,0)[lt]{\lineheight{1.25}\smash{\begin{tabular}[t]{l}$f(\mathbf{x}-\mathbf{e}_2)$\end{tabular}}}}%
	\put(0.02363697,0.18034993){\makebox(0,0)[lt]{\lineheight{1.25}\smash{\begin{tabular}[t]{l}$f(\mathbf{x}-\mathbf{e}_1)$\end{tabular}}}}%
	\put(0.6776974,0.12158603){\makebox(0,0)[lt]{\lineheight{1.25}\smash{\begin{tabular}[t]{l}$F_1(\mathbf{x})$\end{tabular}}}}%
	\put(0.53034779,0.22483994){\makebox(0,0)[lt]{\lineheight{1.25}\smash{\begin{tabular}[t]{l}$-F_2(\mathbf{x})$\end{tabular}}}}%
	\put(0.66055649,0.32686341){\makebox(0,0)[lt]{\lineheight{1.25}\smash{\begin{tabular}[t]{l}$-F_1(\mathbf{x}+\mathbf{e}_2)$\end{tabular}}}}%
	\put(0.79321033,0.22483995){\makebox(0,0)[lt]{\lineheight{1.25}\smash{\begin{tabular}[t]{l}$F_2(\mathbf{x}+\mathbf{e}_1)$\end{tabular}}}}%
	\end{picture}%
	\endgroup%
	\caption{Illustration of the lattice Laplacian $\nabla^2f (\mathbf{x})$ of a scalar field $f(\mathbf{x})$ (left) and the lattice curl $\nabla \times \mathbf{F}~(\mathbf{x})$ of a vector field $F_i(\mathbf{x})$ (right). The lattice Laplacian at a site $\mathbf{x}$ involves all adjacent sites. The resulting field resides again on the sites. The lattice curl transforms a vector field, a field on the links, into a field on the plaquettes, where the plaquette is labeled by the site at its bottom left corner.}
	\label{Figure1_lattice}
\end{figure}

We define difference operators - forward
\begin{equation}
\Delta_i^{\left(+\right)}f\left(\mathbf{x}\right) = f\left(\mathbf{x}+\hat{\mathbf{e}}_i\right) -  f\left(\mathbf{x}\right)
\end{equation}
and backward
\begin{equation}
\Delta_i^{\left(-\right)}f\left(\mathbf{x}\right) = f\left(\mathbf{x}\right) -  f\left(\mathbf{x}-\hat{\mathbf{e}}_i\right)
\end{equation}
(acting similarly on vector and pseudovector fields). Out of those, we can construct the lattice versions of the central differential operators  in vector calculus:
\begin{enumerate}
	\item The \emph{gradient} of a scalar field on the lattice sites is a vector field on the links, involving the field's value on the links ends:
	\begin{equation}
	\left(\nabla f\left(\mathbf{x}\right)\right)_i = \Delta^{(+)}_if\left(\mathbf{x}\right) =f\left(\mathbf{x}+\hat{\mathbf{e}}_i\right) -  f\left(\mathbf{x}\right)
	\end{equation}
	\item The \emph{divergence} of a vector field on the links is a scalar field on the lattice sites. Its value on a site involves the values of the vector components of all the links surrounding it:
	\begin{equation}
	\nabla \cdot \mathbf{F}\left(\mathbf{x}\right) = \Delta^{(-)}_i F_i\left(\mathbf{x}\right) = \underset{i}{\sum}\left(F_i\left(\mathbf{x}\right)-F_i\left(\mathbf{x}-\hat{\mathbf{e}}_i\right)\right)
	\end{equation}
	\item The \emph{Laplacian} of a scalar field or a component of another field is given by combining the gradient and the divergence (see Fig.~\ref{Figure1_lattice}):
	\begin{equation}
	\nabla^2 f\left(\mathbf{x}\right) = \Delta^{(-)}_i\Delta^{(+)}_i f\left(\mathbf{x}\right) 
	= \underset{i}{\sum}\left(f\left(\mathbf{x}+\hat{\mathbf{e}}_i\right)+f\left(\mathbf{x}-\hat{\mathbf{e}}_i\right)\right) - 4f\left(\mathbf{x}\right)
	\end{equation}
	\item The \emph{curl} of a vector field gives rise to a pseudovector residing on the plaquettes (dual lattice sites) as illustrated in Fig.~\ref{Figure1_lattice},
	\begin{equation}
	\nabla \times \mathbf{F} \left(\mathbf{x}\right) = \epsilon_{ij}\Delta^{(+)}_iF_j\left(\mathbf{x}\right)
	\end{equation}
	\item The \emph{curl} of a pseudovector field on the plaquettes gives rise to a vector field on the links,
	\begin{equation}
	\left(\nabla \times L \left(\mathbf{x}\right)\right)_i = \epsilon_{ij}\Delta^{(-)}_jL\left(\mathbf{x}\right)
	\end{equation}
	where $\epsilon_{ij}$ is completely antisymmetric.
\end{enumerate}
 
\begin{figure}
	\centering
	\def\svgwidth{\columnwidth}
	\begingroup%
	\makeatletter%
	\providecommand\color[2][]{%
		\errmessage{(Inkscape) Color is used for the text in Inkscape, but the package 'color.sty' is not loaded}%
		\renewcommand\color[2][]{}%
	}%
	\providecommand\transparent[1]{%
		\errmessage{(Inkscape) Transparency is used (non-zero) for the text in Inkscape, but the package 'transparent.sty' is not loaded}%
		\renewcommand\transparent[1]{}%
	}%
	\providecommand\rotatebox[2]{#2}%
	\newcommand*\fsize{\dimexpr\f@size pt\relax}%
	\newcommand*\lineheight[1]{\fontsize{\fsize}{#1\fsize}\selectfont}%
	\ifx\svgwidth\undefined%
	\setlength{\unitlength}{179.91553938bp}%
	\ifx\svgscale\undefined%
	\relax%
	\else%
	\setlength{\unitlength}{\unitlength * \real{\svgscale}}%
	\fi%
	\else%
	\setlength{\unitlength}{\svgwidth}%
	\fi%
	\global\let\svgwidth\undefined%
	\global\let\svgscale\undefined%
	\makeatother%
	\begin{picture}(1,0.37517604)%
	\lineheight{1}%
	\setlength\tabcolsep{0pt}%
	\put(0,0){\includegraphics[width=\unitlength,page=1]{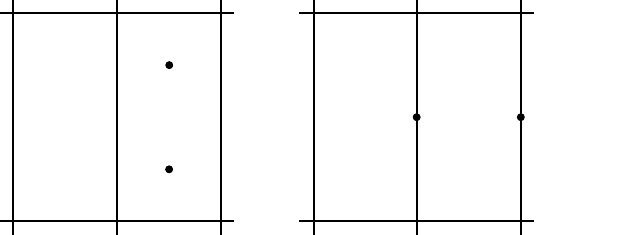}}%
	\put(0.25104578,0.28841209){\makebox(0,0)[lt]{\lineheight{1.25}\smash{\begin{tabular}[t]{l}$L(\mathbf{x})$\end{tabular}}}}%
	\put(0.19421419,0.05502523){\makebox(0,0)[lt]{\lineheight{1.25}\smash{\begin{tabular}[t]{l}$-L(\mathbf{x}-\mathbf{e_2})$\end{tabular}}}}%
	\put(0.58791965,0.20298731){\makebox(0,0)[lt]{\lineheight{1.25}\smash{\begin{tabular}[t]{l}$f(\mathbf{x})$\end{tabular}}}}%
	\put(0.83930247,0.20507161){\makebox(0,0)[lt]{\lineheight{1.25}\smash{\begin{tabular}[t]{l}$-f(\mathbf{x}+\mathbf{e}_1)$\end{tabular}}}}%
	\put(0.70944086,0.20152668){\makebox(0,0)[lt]{\lineheight{1.25}\smash{\begin{tabular}[t]{l}$F_1^L(\mathbf{x})$\end{tabular}}}}%
	\put(0.2262708,0.20152665){\makebox(0,0)[lt]{\lineheight{1.25}\smash{\begin{tabular}[t]{l}$F_1^T(\mathbf{x})$\end{tabular}}}}%
	\put(0,0){\includegraphics[width=\unitlength,page=2]{Figure2_lattice.pdf}}%
	\end{picture}%
	\endgroup%
	\caption{Illustration of the Helmholtz decomposition on the lattice. Analogous to the continuum, a vector field can be split into a transverse component (left) and a longitudinal component (right). The transverse component can be expressed as the lattice curl of a field $L$ on the plaquettes (the analog of a vector potential), whereas the longitudinal component is generated as the (negative) gradient of a scalar field $f$ on the sites. For details on this decomposition see Appendix~\ref{helmholtz}.}
	\label{Figure2_lattice}
\end{figure}

 As in the continuum, each vector field $\mathbf{F}\left(\mathbf{x}\right)$ may be decomposed into the sum of \emph{longitudinal}  and \emph{transverse} parts, $\mathbf{F}^L\left(\mathbf{x}\right)$ and $\mathbf{F}^T\left(\mathbf{x}\right)$ respectively,
 \begin{equation}
\mathbf{F}\left(\mathbf{x}\right) = \mathbf{F}^L\left(\mathbf{x}\right) + \mathbf{F}^T\left(\mathbf{x}\right)
\label{hhd}
 \end{equation}
 The longitudinal part is the gradient of some scalar function $f\left(\mathbf{x}\right)$, and therefore, using the definitions above and similar to the continuous case, it is curl-free:
 \begin{equation}
 \mathbf{F}^L\left(\mathbf{x}\right) = -\nabla f\left(\mathbf{x}\right) \iff \nabla \times  \mathbf{F}^L\left(\mathbf{x}\right) =0
 \label{longdef}
 \end{equation}
 Similarly, the transverse part is the curl of some pseudovector, and its divergence vanishes:
 \begin{equation}
\mathbf{F}^T\left(\mathbf{x}\right) = \nabla \times L\left(\mathbf{x}\right) \iff \nabla \cdot  \mathbf{F}^T\left(\mathbf{x}\right) =0
\label{transdef}
\end{equation} 
 The decomposition into transverse and longitudinal parts (illustrated in Fig.~\ref{Figure2_lattice}), normally referred to as the \emph{Helmholtz decomposition}, is proven similarly to its continuum version, as discussed in Appendix \ref{helmholtz}. This decomposition will be crucial in separating the dynamical (transverse) from the gauge-constrained (longitudinal) degrees of freedom.
 
\subsection{The Matter} \label{mattersection}
Matter particles reside on the lattice sites $\mathbf{x}$. At each site we define an operator $Q\left(\mathbf{x}\right)$ which measures the local charge. It has an integer spectrum, which may be bounded or not, depending on the nature of matter. The charge operators commute with one another,
\begin{equation}
\left[Q\left(\mathbf{x}\right),Q\left(\mathbf{y}\right)\right]=0
\label{Q1}
\end{equation}
We define, on each site, matter field operators $\Psi\left(\mathbf{x}\right)$ which lower the local charge, and their hermitian conjugate which raise it:
\begin{equation}
\begin{aligned}
&\left[Q\left(\mathbf{x}\right),\Psi\left(\mathbf{y}\right)\right]=-\delta\left(\mathbf{x},\mathbf{y}\right)\Psi\left(\mathbf{x}\right)\\
&\left[Q\left(\mathbf{x}\right),\Psi^{\dagger}\left(\mathbf{y}\right)\right]=\delta\left(\mathbf{x},\mathbf{y}\right)\Psi^{\dagger}\left(\mathbf{x}\right)
\label{Q2}
\end{aligned}
\end{equation}
where $\delta\left(\mathbf{x},\mathbf{y}\right)$ is the Kronecker delta function for the lattice discrete coordinates (sites).
	
There are various options to achieve these fairly general commutation relations. In the most common choice, the matter will be fermionic, and each site may host a single species, that is,
\begin{equation}
\left\{\Psi\left(\mathbf{x}\right),\Psi^{\dagger}\left(\mathbf{y}\right)\right\}=\delta\left(\mathbf{x},\mathbf{y}\right)
;
\quad\quad
\left\{\Psi\left(\mathbf{x}\right),\Psi\left(\mathbf{y}\right)\right\}=0
\end{equation}
Then, following Susskind \cite{susskind_lattice_1977}, we can define staggered charges, which split the lattice into two sublattices (even and odd) of particles and anti-particles,
\begin{equation}
Q\left(\mathbf{x}\right)=\left\{
\begin{array}{ll}
\Psi^{\dagger}\left(\mathbf{x}\right)\Psi\left(\mathbf{x}\right), &\mathbf{x}\text{ is even}\\
\Psi^{\dagger}\left(\mathbf{x}\right)\Psi\left(\mathbf{x}\right)-1, &\mathbf{x}\text{ is odd}
\end{array}\right.
\end{equation} 
On even sites, the charges can be $0$ or $1$ while on odd ones they are $-1$ or $0$, depending on whether a fermion is absent or present.
Otherwise, one can use naive or Wilson fermions \cite{zichichi_new_1977}, in which several spin components (two or four) are introduced at each site, and charges are defined with some choice of Dirac matrices implementing the Dirac-Clifford algebra. In all these fermionic options, the desired commutation relations (\ref{Q1}) and (\ref{Q2}) are satisfied.

One could also replace the fermionic matter field by a bosonic field, e.g. a complex scalar field, for which
\begin{equation}
\left[\Psi\left(\mathbf{x}\right),\Psi^{\dagger}\left(\mathbf{y}\right)\right]=
\left[\Psi\left(\mathbf{x}\right),\Psi\left(\mathbf{y}\right)\right]=0
\end{equation}
Each site can host both particles (created by the bosonic mode operator $a^{\dagger}\left(\mathbf{x}\right)$) and  anti-particles (created by the bosonic mode operator $b^{\dagger}\left(\mathbf{x}\right)$), and we expand
\begin{equation}
\Psi\left(\mathbf{x}\right)=\frac{1}{\sqrt{2}}\left(a\left(\mathbf{x}\right)+ib^{\dagger}\left(\mathbf{x}\right)\right)
\end{equation}
The charge operator,
\begin{equation}
Q\left(\mathbf{x}\right)=a^{\dagger}\left(\mathbf{x}\right)a\left(\mathbf{x}\right)-b^{\dagger}\left(\mathbf{x}\right)b\left(\mathbf{x}\right)
\end{equation}
has, in this case, an infinite, non-bounded integer spectrum. The relations (\ref{Q1}) and (\ref{Q2}) are satisfied.

Another way to represent this type of matter field is in the polar representation,
\begin{equation}
\Psi\left(\mathbf{x}\right) = R\left(\mathbf{x}\right)e^{i\varphi\left(\mathbf{x}\right)}
\end{equation}
with two real, commuting scalar fields $R\left(\mathbf{x}\right),\varphi\left(\mathbf{x}\right)$. In the presence of a Higgs potential and following the conventional quasi-classical treatment, the radial degree of freedom is fixed to a constant, $R\left(\mathbf{x}\right)=R_0$ and the remaining compact field $\varphi\left(\mathbf{x}\right)$ is the Goldstone mode \cite{fradkin_phase_1979}. It is canonically conjugate to $Q\left(\mathbf{x}\right)$, that is
\begin{equation}
\left[\varphi\left(\mathbf{x}\right),Q\left(\mathbf{y}\right)\right]=i\delta\left(\mathbf{x},\mathbf{y}\right)
\end{equation}
(the radial field $R\left(\mathbf{x}\right)$ has nothing to do with the charge even if it is not frozen; the reason it may be frozen in a Hamiltonian treatment of the Higgs mechanism is that the Hamiltonian does not contain any terms non-commuting with it, unlike with the angular field).

Typical Hamiltonian terms that involve only the matter will commute with the charge operators. For example, in the case of staggered fermions, one typically uses the mass Hamiltonian \cite{susskind_lattice_1977},
\begin{equation}
H_m = m\underset{\mathbf{x}}{\sum}\left(-1\right)^{x_1+x_2}\Psi^{\dagger}\left(\mathbf{x}\right)\Psi\left(\mathbf{x}\right)
\end{equation}
Or, in the case of the Higgs field, the charge Hamiltonian
\begin{equation}
H_Q = \frac{1}{2R_0^2}\underset{\mathbf{x}}{\sum}Q^2\left(\mathbf{x}\right).
\end{equation}

\subsection{The Gauge Field}

On each link of the lattice we introduce the Hilbert space of a particle on a ring, where the canonical pair of an angular, compact coordinate $\phi_i\left(\mathbf{x}\right)$ and its conjugate $U(1)$ angular momentum operator $E_i\left(\mathbf{x}\right)$, which takes an integer, non-bounded spectrum, is defined, satisfying the canonical relation
\begin{equation}
\left[\phi_i\left(\mathbf{x}\right),E_j\left(\mathbf{y}\right)\right]=i\delta_{ij}\delta\left(\mathbf{x},\mathbf{y}\right).
\label{phiE}
\end{equation} 

$\phi$ plays the role of the (compact) vector potential while $E$ is the electric field. The pure-gauge parts of the Hamiltonian \cite{kogut_hamiltonian_1975,kogut_introduction_1979} are the electric energy term
\begin{equation}
H_E = \frac{g^2}{2}\underset{\mathbf{x},i}{\sum}E_i^2\left(\mathbf{x}\right)= \frac{g^2}{2}\underset{\mathbf{x}}{\sum}E_i\left(\mathbf{x}\right)E_i\left(\mathbf{x}\right)
\end{equation}
(with $g^2$ the coupling constant) and the magnetic energy,
\begin{equation}
H_B = -\frac{1}{g^2}\underset{\mathbf{x}}{\sum}
\cos\left(\phi_1\left(\mathbf{x}\right) + \phi_2\left(\mathbf{x}+\mathbf{e}_1\right)
	- \phi_1\left(\mathbf{x}+\mathbf{e}_2\right)-\phi_2\left(\mathbf{x}\right)\right)
\end{equation}
involving plaquette interactions. The argument of the cosine is nothing but the curl of the vector potential, which is the magnetic field - a pseudovector residing at the center of plaquettes (dual lattice sites):
\begin{equation}
B\left(\mathbf{x}\right)=\nabla \times \mathbf{\phi}\left(\mathbf{x}\right)=\epsilon_{ij}\Delta_i^{\left(+\right)}\phi_j\left(\mathbf{x}\right)
\label{Bdef}
\end{equation}
where $\epsilon_{ij}$ is the completely antisymmetric symbol. Therefore, 
\begin{equation}
H_B = -\frac{1}{g^2}\underset{\mathbf{x}}{\sum}\cos\left(B\left(\mathbf{x}\right)\right)
\end{equation}

The remaining piece of the Hamiltonian couples the matter to the gauge fields. Conventional interaction terms (the result of standard minimal coupling procedures) involve charge hopping to the nearest neighbor site, combined with the increase or decrease of the electric field on the connecting link,
\begin{equation}
H_{int}=\underset{\mathbf{x},i}{\sum} t_{\mathbf{x},i} \Psi^{\dagger}\left(\mathbf{x}\right)e^{i\phi_i\left(\mathbf{x}\right)}
\Psi\left(\mathbf{x}+\hat{\mathbf{e}}_i\right)+h.c.
\end{equation}
with $t_{\mathbf{x},i}$ the tunneling amplitude (which might be position and/or direction dependent). In the case of naive/Wilson fermions, spin components are included, requiring to add some Dirac matrix coupling between them. In all these interactions, the gauge field on the links mediates the movement of the charge to maintain gauge invariance, as shall be discussed now.

\subsection{Gauge Invariance and the Gauss Law}
The full Hamiltonian of a lattice gauge theory as the one discussed above is 
\begin{equation}
H=H_E+H_B+H_{int}+H_m
\label{originalhamiltonian}
\end{equation}
It has a local, or gauge symmetry; that is, there exist local operators $\mathcal{G}\left(\mathbf{x}\right)$, which all commute with the Hamiltonian
\begin{equation}
\left[\mathcal{G}\left(\mathbf{x}\right),H\right]=0\quad\forall \mathbf{x}
\end{equation}
These local symmetry generators are nothing but the \emph{Gauss law operators}, defined by the difference between the electric field divergence on a site and the local charge,
\begin{equation}
\begin{aligned}
\mathcal{G}\left(\mathbf{x}\right) &= \nabla \cdot \mathbf{E}\left(\mathbf{x}\right)-Q\left(\mathbf{x}\right)
=\Delta^{(-)}_iE_i\left(\mathbf{x}\right)-Q\left(\mathbf{x}\right)\\&=\underset{i}{\sum}\left(E_i\left(\mathbf{x}\right)-E_i\left(\mathbf{x}-\hat{\mathbf{e}}_i\right)\right)-Q\left(\mathbf{x}\right)
\end{aligned}
\end{equation}

The commutation  of all the local constants of motion $\mathcal{G}\left(\mathbf{x}\right)$ with the Hamiltonian splits the Hilbert space to different sectors, disconnected by the Hamiltonian dynamics, classified by the eigenstates of these operators $q\left(\mathbf{x}\right)$ which are nothing but static charge configurations, and thus these sectors are simply a formulation of a \emph{charge superselection rule}; so-called \emph{physical states} satisfy
\begin{equation}
\mathcal{G}\left(\mathbf{x}\right)\left|\mathrm{phys}\right\rangle = q\left(\mathbf{x}\right)\left|\mathrm{phys}\right\rangle \quad\forall \mathbf{x}
\end{equation}
or
\begin{equation}
\Delta^{(-)}_iE_i\left(\mathbf{x}\right)\left(\mathbf{x}\right)\left|\mathrm{phys}\right\rangle = \left(Q\left(\mathbf{x}\right)+q\left(\mathbf{x}\right)\right)\left|\mathrm{phys}\right\rangle \quad\forall \mathbf{x}
\label{Gauss}
\end{equation}

Below, we will always assume that the static charges $q\left(\mathbf{x}\right)$ are fixed, which we can do due to the superselection rule.

\section{Decoupling the Matter}\label{decoupsec}
In order to arrive at a redundancy-free formulation, it is important to single out the degrees of freedom which are constrained by the Gauss law constraints. By inspection of eq. (\ref{Gauss}), it becomes clear that the longitudinal part of the electric field is completely determined by the charge configuration in the physical Hilbert space. The divergence-free (transverse) part of the electric field is not affected by these constraints. 

Compared to previous references where only static charges were considered, it is not as straightforward in the presence of dynamical matter to write down a Hamiltonian in terms of transverse gauge field degrees of freedom. This is due to the appearance of gauge-matter interactions $H_{int}$, which involve the longitudinal component of the gauge field, whereas in the magnetic Hamiltonian $H_{B}$ only the transverse component contributes. 

The idea is to find a unitary transformation to a frame in which the longitudinal part of the gauge field disappears from the Hamiltonian (one can intuitively think about it as rotating to a frame such that Coulomb gauge holds in the physical subspace, $ \Delta^{(-)}_i \phi_i \left( \mathbf{x} \right)=0$). A unitary transformation $\mathcal{U}$, which accomplishes that, can be defined as 
\begin{equation}
\mathcal{U}=\exp\left(-i\underset{\mathbf{x}}{\sum}\phi_i \left(\mathbf{x}\right)\beta_i \left(\mathbf{x}\right)\right)
\end{equation}
with 
\begin{equation}
\beta_i \left(\mathbf{x}\right) = - \underset{\mathbf{y}}{\sum}\Delta^{(+)}_{i,\mathbf{x}}G\left(\mathbf{x},\mathbf{y}\right)\left(Q\left(\mathbf{y}\right)+q\left(\mathbf{y}\right)\right)
\label{beta}
\end{equation}

where $G\left(\mathbf{x},\mathbf{y}\right)$ is the Green's function of the (negative) lattice Laplacian (see Appendix \ref{greensfunction}). $\Delta^{(+)}_{i,\mathbf{x}}$ denotes the lattice forward derivative in direction $\mathbf{\hat{e}}_i$ with respect to the variable $\mathbf{x}$. $\beta_{i} (\mathbf{x})$ is nothing but the longitudinal electric field in the physical Hilbert space before the transformation, $\beta_{i} (\mathbf{x})\left|\mathrm{phys}\right\rangle = E^L_{i} (\mathbf{x})\left|\mathrm{phys}\right\rangle$ (see Appendix \ref{helmholtz} for details).

 We start by studying the effect of this transformation on the Gauss law in eq. (\ref{Gauss}). It is clear that the charge operator $Q(\mathbf{x})$ commutes with the transformation: 
\begin{equation}
\mathcal{U} Q\left(\mathbf{x}\right)\mathcal{U}^{\dagger} = Q\left(\mathbf{x}\right).
\end{equation}
The electric field gets shifted by $\beta$:
\begin{equation}
\mathcal{U}E_i \left(\mathbf{x}\right)\mathcal{U}^{\dagger}=E_i \left(\mathbf{x}\right)+\beta_i \left(\mathbf{x}\right)
\label{Etrans}
\end{equation}
Thus, the divergence of the electric field gives
\begin{equation}
\begin{aligned}
\mathcal{U} \Delta^{(-)}_iE_i\left(\mathbf{x}\right)\mathcal{U}^{\dagger} &=\Delta^{(-)}_iE_i\left(\mathbf{x}\right) - \sum_{\mathbf{y}}
\Delta_{i,\mathbf{x}}^{(-)}\Delta^{(+)}_{i,\mathbf{x}}G\left(\mathbf{x},\mathbf{y}\right)\left(Q\left(\mathbf{y}\right)+q\left(\mathbf{y}\right)\right) \\
&= \Delta^{(-)}_iE_i\left(\mathbf{x}\right)+ Q\left(\mathbf{x}\right)+q\left(\mathbf{x}\right)
\end{aligned}
\end{equation}

Hence, the physical states in the rotated frame, $ \left| \widetilde{\mathrm{phys}}\right\rangle \equiv \mathcal{U} \left| \mathrm{phys}\right\rangle $, obey the transformed matter-free Gauss law,
\begin{equation}
\Delta^{(-)}_iE_i\left(\mathbf{x}\right)\left|\widetilde{\mathrm{phys}}\right\rangle=0\quad\forall \mathbf{x}
\label{GaussT}
\end{equation}
or, in other words, the electric field in the physical subspace is transverse (divergence-free) after the unitary transformation $\mathcal{U}$. This was to be expected since we removed the longitudinal part of the electric field (in the physical subspace). 

Note that the spectrum of $E_i\left(\mathbf{x}\right)$ has changed in the rotated physical subspace: the integer spectrum becomes fractional after the transformation. As shown in \cite{drell_quantum_1979,PhysRevResearch.2.043145}, every charge configuration introduces in the physical Hilbert space a certain constant shift in the original integer spectrum of the electric field. However, in the presence of dynamical matter, the different static charge sectors get mixed due to the gauge-matter interactions. Therefore, the spectrum is a shifted integer spectrum where the shift is not fixed but depends on the charge configuration. 

In the next step, we consider the transformation of the electric part of the Hamiltonian. Using (\ref{Etrans}) and (\ref{beta}), we obtain that the transformed electric Hamiltonian has three parts,
\begin{equation}
\begin{aligned}
\tilde H_E &= \mathcal{U} H_E \mathcal{U}^{\dagger} \\
&= \frac{g^2}{2} \sum_{\mathbf{x},i} \left( E_{i}(\mathbf{x}) - \underset{\mathbf{y}}{\sum}\Delta^{(+)}_{i,\mathbf{x}}G\left(\mathbf{x},\mathbf{y}\right)\left(Q\left(\mathbf{y}\right)+q\left(\mathbf{y}\right)\right) \right)^2 \\
&\equiv \tilde H^T_E + \tilde H^L_E + \tilde H^{TL}_E
\end{aligned}
\end{equation}
The first term will have the same mathematical form as the pre-transformed Hamiltonian, but now, in the physical Hilbert space, it will only correspond to the transverse parts of the field (which no longer possess an integer spectrum),
\begin{equation}
\tilde H^T_E = \frac{g^2}{2}\underset{\mathbf{x},i}{\sum}E_i^2\left(\mathbf{x}\right)= \frac{g^2}{2}\underset{\mathbf{x}}{\sum}E_i\left(\mathbf{x}\right)E_i\left(\mathbf{x}\right)
\label{HT}
\end{equation}
The second part is the purely longitudinal one, taking the form
\begin{equation}
\begin{aligned}
\tilde H^L_E &=- \frac{g^2}{2} \underset{\mathbf{x},\mathbf{y},\mathbf{y'},i}{\sum}\Delta^{(-)}_{i,\mathbf{x}} \Delta^{(+)}_{i,\mathbf{x}}G\left(\mathbf{x},\mathbf{y}\right)\left(Q\left(\mathbf{y}\right)+q\left(\mathbf{y}\right)\right) \times \\
& \hspace{45pt} \times G\left(\mathbf{x},\mathbf{y'}\right)\left(Q\left(\mathbf{y'}\right)+q\left(\mathbf{y'}\right)\right) \\ 
&=\frac{g^2}{2}\underset{\mathbf{x},\mathbf{y}}{\sum}
\left(Q\left(\mathbf{x}\right)+q\left(\mathbf{x}\right)\right)
G\left(\mathbf{x},\mathbf{y}\right)
\left(Q\left(\mathbf{y}\right)+q\left(\mathbf{y}\right)\right)
\end{aligned}
\label{HL}
\end{equation}
where in the first row we used a lattice analogue of integrating by parts (which is valid for both periodic and open boundary conditions) and in the second row the definition of the Green's function. The resulting interaction between the charges (both dynamical and static) is of Coulomb-type, since the Green's function $G(\mathbf{x},\mathbf{y})$ is nothing but the lattice Coulomb potential generated by a unit charge at $\mathbf{y}$ experienced by another unit charge at $\mathbf{x}$. 

The third part involves the cross terms. If we write it as  
\begin{equation}
\begin{aligned}
\tilde{H}_{E}^{TL}&=-g^2 \sum_{\mathbf{x},\mathbf{y},i} E_{i}(\mathbf{x}) \Delta^{(+)}_{i,\mathbf{x}}G\left(\mathbf{x},\mathbf{y}\right)\left(Q\left(\mathbf{y}\right)+q\left(\mathbf{y}\right)\right) \\
&= g^2 \sum_{\mathbf{x},\mathbf{y},i} \Delta^{(-)}_{i,\mathbf{x}} E_{i}(\mathbf{x}) G\left(\mathbf{x},\mathbf{y}\right)\left(Q\left(\mathbf{y}\right)+q\left(\mathbf{y}\right)\right)
\end{aligned}
\end{equation}
it becomes clear that it vanishes in the physical Hilbert space using the transformed Gauss law in eq. (\ref{GaussT}). Intuitively, it can be understood since it corresponds to the scalar product of the longitudinal and transverse component of the electric field. We are only interested in the physical subspace and will therefore neglect this term in the following.

To study the transformation of the matter degrees of freedom it is useful to rewrite the transformation $\mathcal{U}$ in the following way: 
\begin{equation}
\begin{aligned}
\mathcal{U}&=\exp\left(i\underset{\mathbf{x},\mathbf{y}}{\sum}\phi_i \left(\mathbf{x}\right)\Delta^{(+)}_{i,\mathbf{x}}G\left(\mathbf{x},\mathbf{y}\right)\left(Q\left(\mathbf{y}\right)+q\left(\mathbf{y}\right)\right)\right)\\
&=\exp\left(-i\underset{\mathbf{x},\mathbf{y}}{\sum}
\left(Q\left(\mathbf{x}\right)+q\left(\mathbf{x}\right)\right)
G\left(\mathbf{x},\mathbf{y}\right)
\Delta^{(-)}_{i,\mathbf{y}}\phi_i \left(\mathbf{y}\right)\right)
\end{aligned}
\end{equation}
(where we used again the lattice analog of integrating by parts). Using (\ref{Q1}) and (\ref{Q2}), we obtain the transformation rule of the charge raising operator,
\begin{equation}
\mathcal{U} \Psi^{\dagger}\left(\mathbf{x}\right) \mathcal{U}^{\dagger}=
\Psi^{\dagger}\left(\mathbf{x}\right)
\exp\left(-i\underset{\mathbf{y}}{\sum}
G\left(\mathbf{x},\mathbf{y}\right)
\Delta^{(-)}_{i,\mathbf{y}}\phi_i \left(\mathbf{y}\right)\right)
\end{equation}
and thus the gauge-matter interactions in the transformed picture are
\begin{widetext}
\begin{equation}
\tilde H_{int}=\mathcal{U}H_{int}\mathcal{U}^{\dagger}=\underset{\mathbf{x},i}{\sum} t_{\mathbf{x},i}\Psi^{\dagger}\left(\mathbf{x}\right)\exp\left[i\left(\phi_i\left(\mathbf{x}\right)+\underset{\mathbf{y}}{\sum}
\Delta^{(+)}_{i,\mathbf{x}}G\left(\mathbf{x},\mathbf{y}\right)
\Delta^{(-)}_{i,\mathbf{y}}\phi_i \left(\mathbf{y}\right)\right)\right]
\Psi\left(\mathbf{x}+\hat{\mathbf{e}}_i\right)+h.c.
\label{transformedgaugematter}
\end{equation}
\end{widetext}

Using the Helmholtz decomposition (see Appendix \ref{helmholtz} for details), one obtains that the longitudinal part of $\phi_i\left(\mathbf{x}\right)$ is given by
\begin{equation}
\phi^L_i\left(\mathbf{x}\right)=-\Delta^{(+)}_{i,\mathbf{x}} \sum_{\mathbf{y}} G\left(\mathbf{x},\mathbf{y}\right)
\Delta^{(-)}_{i,\mathbf{y}}\phi_i \left(\mathbf{y}\right)
\end{equation}
and hence we remain only with the transverse, divergence free field in the transformed interaction Hamiltonian,
\begin{equation}
\tilde H_{int} =\underset{\mathbf{x},i}{\sum} t_{\mathbf{x},i} \left(\Psi^{\dagger}\left(\mathbf{x}\right)e^{i\phi^{T}_i\left(\mathbf{x}\right)}
\Psi\left(\mathbf{x}+\hat{\mathbf{e}}_i\right)+h.c.\right)
\label{Hinttilde}
\end{equation}
This can also easily be checked by taking the lattice divergence of the argument in the exponential in eq. (\ref{transformedgaugematter}).

$H_B$ does not change under the transformation, because it commutes with $\mathcal{U}$, i.e. $\tilde{H}_B=H_B$. It depends on the \emph{curl} of the vector potential $\phi $ so that only the transverse part of $\phi$ contributes (since the longitudinal one is curl-free). Therefore, $\tilde H_B$ can be formulated with the transverse field only,
\begin{equation}
\tilde H_B = -\frac{1}{g^2}\underset{\mathbf{x}}{\sum}\cos\left(\epsilon_{ij}\Delta_i^{\left(+\right)}\phi^T_j\left(\mathbf{x}\right)\right)
\end{equation}
$H_m$ commutes with $\mathcal{U}$ as well, and thus $\tilde H_m = H_m$. 

Hence, after the transformation, the Hamiltonian depends only on the transverse component of the vector potential, $\phi^T$, so that we indeed transformed to a frame where the lattice version of Coulomb gauge holds. We can therefore proceed to formulate the remaining transverse degrees of freedom in terms of dual variables. For that, we will restrict ourselves from now on to the physical Hilbert space. 

\section{Dual Formulation} \label{dualformulation}

In the transformed picture, the Gauss law (\ref{Gauss}) becomes decoupled from the matter degrees of freedom, leaving the electric field transverse (\ref{GaussT}). Therefore it makes sense to change from the gauge field variables on the links to another set of variables that will respect this transverse nature of the electric field. This will allow us to directly incorporate the Gauss law constraint (\ref{GaussT}), making the formulation manifestly gauge-invariant.

\begin{figure}
	\centering
	\def\svgwidth{\columnwidth}
	\begingroup%
	\makeatletter%
	\providecommand\color[2][]{%
		\errmessage{(Inkscape) Color is used for the text in Inkscape, but the package 'color.sty' is not loaded}%
		\renewcommand\color[2][]{}%
	}%
	\providecommand\transparent[1]{%
		\errmessage{(Inkscape) Transparency is used (non-zero) for the text in Inkscape, but the package 'transparent.sty' is not loaded}%
		\renewcommand\transparent[1]{}%
	}%
	\providecommand\rotatebox[2]{#2}%
	\newcommand*\fsize{\dimexpr\f@size pt\relax}%
	\newcommand*\lineheight[1]{\fontsize{\fsize}{#1\fsize}\selectfont}%
	\ifx\svgwidth\undefined%
	\setlength{\unitlength}{238.16282108bp}%
	\ifx\svgscale\undefined%
	\relax%
	\else%
	\setlength{\unitlength}{\unitlength * \real{\svgscale}}%
	\fi%
	\else%
	\setlength{\unitlength}{\svgwidth}%
	\fi%
	\global\let\svgwidth\undefined%
	\global\let\svgscale\undefined%
	\makeatother%
	\begin{picture}(1,0.86064412)%
	\lineheight{1}%
	\setlength\tabcolsep{0pt}%
	\put(0,0){\includegraphics[width=\unitlength,page=1]{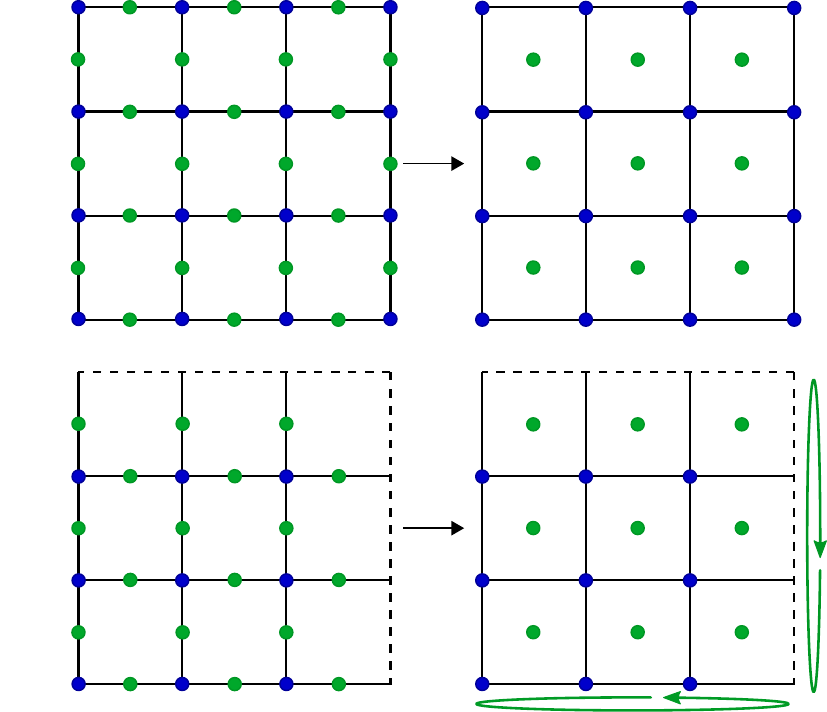}}%
	\put(-0.00082749,0.65039121){\makebox(0,0)[lt]{\lineheight{1.25}\smash{\begin{tabular}[t]{l}$\mathbf{OBC}$\end{tabular}}}}%
	\put(-0.00093229,0.20974164){\makebox(0,0)[lt]{\lineheight{1.25}\smash{\begin{tabular}[t]{l}$\mathbf{PBC}$\end{tabular}}}}%
	\end{picture}%
	\endgroup%
	\caption{Illustration of the dual formulation in the transformed frame for a $3 \times 3$ lattice with both open boundary conditions (upper row) and periodic boundary conditions (lower row). In the left column, the original formulation is shown in terms of the matter degrees of freedom on the sites (blue) and the gauge degrees of freedom on the links (green). In the right column, the degrees of freedom of the dual formulation are shown: the matter still resides on the lattice sites, but the gauge degrees of freedom are described in terms of plaquette variables. While in the original formulation there were gauge constraints for every site, there are no local gauge constraints left in the dual formulation. For open boundary conditions, there are no gauge constraints in the dual formulation and for periodic boundary conditions there is a global constraint left involving all plaquette variables. Since periodic boundary conditions allow closed loops around the lattice, there are two global gauge variables, one for each spatial direction (green circles in the figure). }
	\label{dualvariables_figure}
\end{figure}

Since the electric field in the physical Hilbert space of the transformed frame is transverse, we may express it as the curl of a pseudovector field $L\left(\mathbf{x}\right)$ defined on the plaquettes (dual lattice sites), $\nabla \times L\left(\mathbf{x}\right)$. If we apply the lattice curl again, this gives rise to a Poisson equation for $L\left(\mathbf{x}\right)$ in terms of $E_i(\mathbf{x})$, whose solution is
\begin{equation}
L\left(\mathbf{x}\right)=\underset{\mathbf{y}}{\sum}G\left(\mathbf{x},\mathbf{y}\right)\epsilon_{ij}\Delta^{(+)}_{i,\mathbf{y}}E_j\left(\mathbf{y}\right)
\label{LE}
\end{equation}
Using that, one can show that $L\left(\mathbf{x}\right)$ is canonically conjugate to the magnetic field (see Appendix~\ref{cancomrel} for details),
\begin{equation}
\left[B\left(\mathbf{x}\right),L\left(\mathbf{y}\right)\right]=i\delta(\mathbf{x},\mathbf{y})
\label{comrelBL}
\end{equation}
The idea behind these variables is that all transverse configurations of the electric field are made out of loops and the local $B$/$L$-variables are a good basis to construct these loops. This geometric picture is also the basis for the dual formulation in \cite{unmuthyockey2019gaugeinvariant}. However, with periodic boundary conditions there are two global loops (up to modifications by plaquette operators, similar to the toric code) that can not be created out of the $B$/$L$-variables (they do not appear in the case of open boundary conditions). These are independent variables, denoted as $B_1,L_1$ and $B_2,L_2$. We choose the $B_1$/$L_1$-variable to wind around the torus along the $\mathbf{\hat{e}}_1$-axis whereas the $B_2$/$L_2$-variable is chosen to wind around the torus along the $\mathbf{\hat{e}}_2$-axis. The sets of degrees of freedom in the dual formulation for both periodic and open boundary conditions are illustrated in Fig.~\ref{dualvariables_figure}, exemplary for a $3 \times 3$ lattice. Therefore, to express the electric field in terms of $L$-variables for periodic boundary conditions, we need in addition to the curl of $L$ the contributions of the global loops, i.e. 
\begin{equation}
E_i\left(\mathbf{x}\right)=\epsilon_{ij} \Delta^{(-)}_jL\left(\mathbf{x}\right)  + \delta_{x_j,0} L_i
\label{EL}
\end{equation}
with $i \neq j$. The second term is only present on the two axes and vanishes completely for open boundary conditions. By formulating the theory in terms of dual variables, there are no local constraints left. However, there is a global constraint left (in case of periodic boundary conditions, this is not the case for open boundary conditions) which can be seen by summing eq.~(\ref{Bdef}) over the whole lattice: 
\begin{equation}
\sum_{\mathbf{x}} B({\mathbf{x}}) \left|\mathrm{phys}\right\rangle = 0 
\label{globalconstraintBL}
\end{equation}
This is intuitively clear since raising the electric flux around every plaquette should give the same state (on a lattice with periodic boundary conditions). To convince us that the number of physical degrees of freedom in the dual formulation matches the number in the original formulation we can do a short counting of degrees of freedom. We can neglect the matter degrees of freedom in this calculation since it is the same in both cases. 

Starting with periodic boundary conditions, there are in the original formulation $2N^2$ links and $N^2-1$ Gauss laws (the constraint at one lattice site is redundant). The number of physical degrees of freedom is thus $N^2+1$. In the dual formulation, we have $N^2$ plaquette variables, two global loop variables and one global constraint, which amounts also to $N^2+1$ physical degrees of freedom. 
 
With open boundary conditions, there are originally $2N(N+1)$ links and $(N+1)^2-1$ Gauss law constraints, i.e. $N^2$ physical degrees of freedom. In the dual formulation, there are $N^2$ plaquette variables which are not subject to any constraints, thus giving the same number of physical degrees of freedom.

We can now rewrite the transformed Hamiltonian in terms of the dual variables. This does not change $\tilde H_m$ and $\tilde H_E^L$; the magnetic part will now be non-interacting,
\begin{equation}
\tilde H_B = -\frac{1}{g^2}\underset{\mathbf{x}}{\sum}\cos\left(B\left(\mathbf{x}\right)\right).
\label{magnetichamiltonianintermsofB}
\end{equation}
Following eq.~(\ref{EL}), the transverse electric part will involve some local interactions,
\begin{equation}
\tilde H^T_E = \frac{g^2}{2}\underset{\mathbf{x},i}{\sum} \left( \epsilon_{ij} \left(L\left(\mathbf{x}\right)- L\left(\mathbf{x}-\mathbf{\hat{e}}_j\right) \right)+ \delta_{x_j,0} L_i \right)^2
\label{etransL}
\end{equation}
where the last term denotes the contribution of the two global loops which is only present on the axes ($x_1=0$ or $x_2=0$). This term drops out in the case of open boundary conditions. 

To rewrite the gauge-matter interactions in terms of dual variables, we express the transverse part of the gauge field in terms of the magnetic field $B(\mathbf{x})$ (the calculation of the shifts $s_{\mathbf{x},i}(\mathbf{y})$ is presented in Appendix \ref{helmholtz})
\begin{equation}
\begin{aligned}
\phi_{i}^T(\mathbf{x})&=\sum_{\mathbf{y}} s_{\mathbf{x},i}(\mathbf{y}) B(\mathbf{y})
\end{aligned}
\label{expressphiTwithshifts}
\end{equation}
(note that the sum over $\mathbf{y}$ also contains the two global loops $B_1$ and $B_2$). The gauge-matter interactions then take the form
\begin{equation}
\tilde H_{int} =\underset{\mathbf{x},i}{\sum} t_{\mathbf{x},i} \Psi^{\dagger}\left(\mathbf{x}\right)e^{i \sum_{\mathbf{y}} s_{\mathbf{x},i}(\mathbf{y}) B(\mathbf{y})}
\Psi\left(\mathbf{x}+\hat{\mathbf{e}}_i\right)+h.c.
\label{dualgaugematter2d}
\end{equation}

The hopping of a matter degree of freedom from some site $\mathbf{x}$ to an adjacent site $\mathbf{x}+\mathbf{\hat{e}}_i$ introduces shifts $s_{\mathbf{x},i}(\mathbf{y})$ ($-1/2 < s_{\mathbf{x},i}(\mathbf{y}) \leq 1/2$) in the $L(\mathbf{y})$ operators since $B$ is canonically conjugate to $L$. This can be understood in the following way: the hopping changes the electric field configuration (by raising/lowering the electric field on that link) and the change in the transverse part of the electric field is characterized by the $s$-shifts. Although the size of these shifts decays with distance to the link where hopping occurs, the interaction involves many degrees of freedom which might be difficult to deal with, in particular for a quantum simulation. Summing up, the whole transformed Hamiltonian in the dual formulation with $B/L$-variables takes the form 
\begin{widetext}
\begin{equation}	
\begin{aligned}
\tilde{H} =& H_m 
+\frac{g^2}{2}\underset{\mathbf{x},i}{\sum} \left( \epsilon_{ij} \left(L\left(\mathbf{x}\right)- L\left(\mathbf{x}-\mathbf{\hat{e}}_j\right)\right)+ \delta_{x_j,0} L_i \right)^2
+\frac{g^2}{2}\underset{\mathbf{x},\mathbf{y}}{\sum} \left(Q\left(\mathbf{x}\right)+q\left(\mathbf{x}\right)\right)
G\left(\mathbf{x},\mathbf{y}\right)
\left(Q\left(\mathbf{y}\right)+q\left(\mathbf{y}\right)\right) \\
&-\frac{1}{g^2}\underset{\mathbf{x}}{\sum}\cos\left(B\left(\mathbf{x}\right)\right)
+ \underset{\mathbf{x},i}{\sum} t_{\mathbf{x},i}\Psi^{\dagger}\left(\mathbf{x}\right)e^{i \sum_{\mathbf{y}} s_{\mathbf{x},i}\left( \mathbf{y} \right) B\left(\mathbf{y}\right)}
\Psi\left(\mathbf{x}+\hat{\mathbf{e}}_i\right)+h.c.
\end{aligned}
\end{equation}
\end{widetext}
with $i \neq j$. In the case of open boundary conditions, the global loop contributions in $\tilde{H}_E^T$ and $\tilde{H}_{int}$ vanish. 

We proceed to define another canonical pair of operators which makes the gauge-matter interactions local again. The idea is to carry out the same procedure as before but now with the transverse component of the gauge field $\phi_{i}^T(\mathbf{x})$ as a starting point, instead of the electric field.

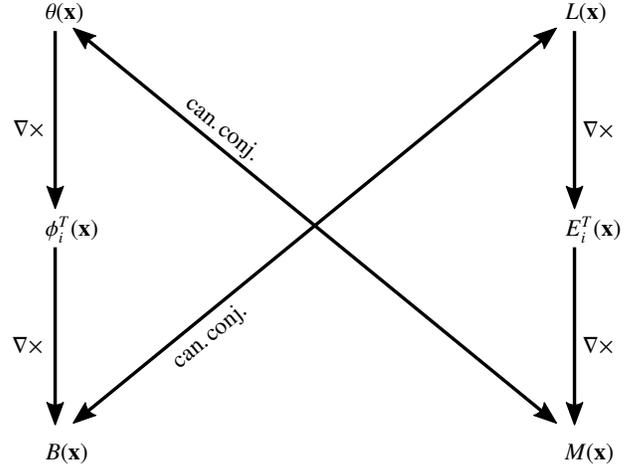
\begin{figure}
	\centering
	\def\svgwidth{\columnwidth}
	\begingroup%
	\makeatletter%
	\providecommand\color[2][]{%
		\errmessage{(Inkscape) Color is used for the text in Inkscape, but the package 'color.sty' is not loaded}%
		\renewcommand\color[2][]{}%
	}%
	\providecommand\transparent[1]{%
		\errmessage{(Inkscape) Transparency is used (non-zero) for the text in Inkscape, but the package 'transparent.sty' is not loaded}%
		\renewcommand\transparent[1]{}%
	}%
	\providecommand\rotatebox[2]{#2}%
	\newcommand*\fsize{\dimexpr\f@size pt\relax}%
	\newcommand*\lineheight[1]{\fontsize{\fsize}{#1\fsize}\selectfont}%
	\ifx\svgwidth\undefined%
	\setlength{\unitlength}{112.70489721bp}%
	\ifx\svgscale\undefined%
	\relax%
	\else%
	\setlength{\unitlength}{\unitlength * \real{\svgscale}}%
	\fi%
	\else%
	\setlength{\unitlength}{\svgwidth}%
	\fi%
	\global\let\svgwidth\undefined%
	\global\let\svgscale\undefined%
	\makeatother%
	\begin{picture}(1,0.7639318)%
	\lineheight{1}%
	\setlength\tabcolsep{0pt}%
	\put(0.08639932,0.69565339){\makebox(0,0)[lt]{\lineheight{1.25}\smash{\begin{tabular}[t]{l}$\theta(\mathbf{x})$\end{tabular}}}}%
	\put(0.88540754,0.69565341){\makebox(0,0)[lt]{\lineheight{1.25}\smash{\begin{tabular}[t]{l}$L(\mathbf{x})$\end{tabular}}}}%
	\put(0.88388804,0.0178494){\makebox(0,0)[lt]{\lineheight{1.25}\smash{\begin{tabular}[t]{l}$M(\mathbf{x})$\end{tabular}}}}%
	\put(0.08637786,0.01784937){\makebox(0,0)[lt]{\lineheight{1.25}\smash{\begin{tabular}[t]{l}$B(\mathbf{x})$\end{tabular}}}}%
	\put(0.08663816,0.36292598){\makebox(0,0)[lt]{\lineheight{1.25}\smash{\begin{tabular}[t]{l}$\phi_i^T(\mathbf{x})$\end{tabular}}}}%
	\put(0.88568951,0.36292622){\makebox(0,0)[lt]{\lineheight{1.25}\smash{\begin{tabular}[t]{l}$E_i^T(\mathbf{x})$\end{tabular}}}}%
	\put(0.00043462,0.51555274){\makebox(0,0)[lt]{\lineheight{1.25}\smash{\begin{tabular}[t]{l}$\hspace{9pt} \nabla\times$\end{tabular}}}}%
	\put(-0.00110399,0.18282503){\makebox(0,0)[lt]{\lineheight{1.25}\smash{\begin{tabular}[t]{l}$\hspace{9pt} \nabla \times$\end{tabular}}}}%
	\put(0.91266945,0.51444777){\makebox(0,0)[lt]{\lineheight{1.25}\smash{\begin{tabular}[t]{l}$\nabla\times$\end{tabular}}}}%
	\put(0.91275462,0.1817205){\makebox(0,0)[lt]{\lineheight{1.25}\smash{\begin{tabular}[t]{l}$\nabla \times$\end{tabular}}}}%
	\put(0,0){\includegraphics[width=\unitlength,page=1]{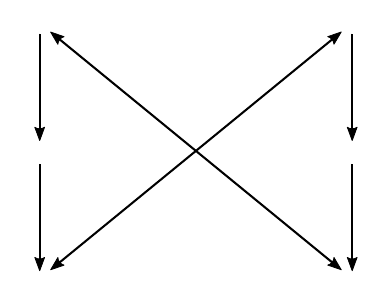}}%
	\put(0.29266528,0.15818037){\rotatebox{39.28900022}{\makebox(0,0)[lt]{\lineheight{1.25}\smash{\begin{tabular}[t]{l}$\mathrm{can.} \hspace{1pt} \mathrm{conj.}$\end{tabular}}}}}%
	\put(0.30051611,0.56310536){\rotatebox{-39.28900022}{\makebox(0,0)[lt]{\lineheight{1.25}\smash{\begin{tabular}[t]{l}$\mathrm{can.} \hspace{1pt} \mathrm{conj.}$\end{tabular}}}}}%
	\end{picture}%
	\endgroup%
	\caption{Illustration of the two dual formulations in terms of $B/L$-variables and $\theta/M$-variables. Both formulations are based on expressing the transverse part of either the gauge field $\phi_i(\mathbf{x})$ or the electric field $E_i(\mathbf{x})$. While the transverse component of the electric field, $E_i^T(\mathbf{x})$, can be obtained as the lattice curl $\nabla \times$ of the plaquette field $L$, the lattice curl of the plaquette field $\theta$ gives rise to $\phi^T_i(\mathbf{x})$. It can then be shown that the curl of $\phi$, which is the magnetic field $B$, is canonically conjugate to $L$. In the same way, it can be shown that the curl of $E$ is canonically conjugate to $\theta$. Thus, the two dual formulations are based on the same principle and complement each other.}
	\label{daul_variables_sketch_figure}
\end{figure}

In complete analogy to eq.~(\ref{EL}), we can express  $\phi_{i}^T(\mathbf{x})$ by a compact field on the plaquettes $\theta\left(\mathbf{x}\right)$ (and for periodic boundary conditions two additional global loops $\theta_1$ and $\theta_2$), such that  
\begin{equation}
\phi^T_i\left(\mathbf{x}\right)= \epsilon_{ij}  \Delta^{(-)}_j\theta\left(\mathbf{x}\right) +\delta_{x_j,0} \theta_i 
\end{equation}
with $i \neq j$. As before, the global loop contribution vanishes for open boundary conditions. The expression for $\theta$ in terms of $\phi$ has the same form as the expression for $L$ in terms of $E$ in eq.~(\ref{LE}):
\begin{equation}
\begin{aligned}
\theta\left(\mathbf{x}\right)=\underset{\mathbf{y}}{\sum}G\left(\mathbf{x},\mathbf{y}\right)\epsilon_{ij}\Delta^{(+)}_{i,\mathbf{y}}\phi_j\left(\mathbf{y}\right)
\end{aligned}
\label{tintermsofb}
\end{equation}
The canonically conjugate variable to $\theta$ is the curl of the electric field, 
\begin{equation}
M\left(\mathbf{x}\right)=\epsilon_{ij}\Delta^{(+)}_i E_j\left(\mathbf{x}\right)
\label{ME}
\end{equation}
Since $E_{i}(\mathbf{x})$ is integer-valued, $M(\mathbf{x})$, as the sum of integer-valued operators, will also have an integer spectrum. Using the expression for $\theta$ in terms of $\phi$ from eq.~(\ref{tintermsofb}), the definition of $M$ in terms of $E$ in eq.~(\ref{ME}), one can show that also $\theta$ and $M$ fulfill the canonical commutation relations,
\begin{equation}
\left[\theta\left(\mathbf{x}\right),M\left(\mathbf{y}\right)\right]=i\delta(\mathbf{x},\mathbf{y}).
\end{equation} 
Analogous to the $B$/$L$-variables, there are two global non-contractible loops denoted as $\theta_1/M_1$ and $\theta_2/M_2$ winding along the respective axis (again, this is only the case for periodic boundary conditions, not for open ones). Since both dual formulations share the same locations on the lattice, the counting of degrees of freedom can be done in the same way as for the $B$/$L$-variables. The relation between the two sets of dual variables is illustrated in Fig.~\ref{daul_variables_sketch_figure}. 

If we express the gauge-matter interactions in terms of the newly introduced field $\theta\left(\mathbf{x}\right)$, we arrive at
\begin{equation}
\tilde H_{int} =\underset{\mathbf{x},i}{\sum} t_{\mathbf{x},i}  \Psi^{\dagger}\left(\mathbf{x}\right)e^{i  \left( \epsilon_{ij}\Delta^{(-)}_j\theta\left(\mathbf{x}\right) + \delta_{x_j,0} \theta_i \right)} \Psi\left(\mathbf{x}+\hat{\mathbf{e}}_i\right)+h.c.
\end{equation}
with $i \neq j$, a local interaction again (up to the contribution of the global loops $\theta_i$ which is only present on the two axes and vanishes in the case of open boundary conditions). It has the following interpretation: when a matter charge hops from site $\mathbf{x}$ to a neighboring site, say $\mathbf{x}+\mathbf{\hat{e}}_{1}$, the curl of the electric field on the plaquette above the link gets raised by one and the curl of the electric field on the plaquette below the link gets lowered by one. In the case of bosonic matter as discussed in Sec.~\ref{mattersection}, the gauge-matter interactions become completely symmetric between the matter and gauge field degrees of freedom. 
\begin{figure*}
	\centering
	\def\svgwidth{\textwidth}
	\begingroup%
	\makeatletter%
	\providecommand\color[2][]{%
		\errmessage{(Inkscape) Color is used for the text in Inkscape, but the package 'color.sty' is not loaded}%
		\renewcommand\color[2][]{}%
	}%
	\providecommand\transparent[1]{%
		\errmessage{(Inkscape) Transparency is used (non-zero) for the text in Inkscape, but the package 'transparent.sty' is not loaded}%
		\renewcommand\transparent[1]{}%
	}%
	\providecommand\rotatebox[2]{#2}%
	\newcommand*\fsize{\dimexpr\f@size pt\relax}%
	\newcommand*\lineheight[1]{\fontsize{\fsize}{#1\fsize}\selectfont}%
	\ifx\svgwidth\undefined%
	\setlength{\unitlength}{455.97312695bp}%
	\ifx\svgscale\undefined%
	\relax%
	\else%
	\setlength{\unitlength}{\unitlength * \real{\svgscale}}%
	\fi%
	\else%
	\setlength{\unitlength}{\svgwidth}%
	\fi%
	\global\let\svgwidth\undefined%
	\global\let\svgscale\undefined%
	\makeatother%
	\begin{picture}(1,0.81641234)%
	\lineheight{1}%
	\setlength\tabcolsep{0pt}%
	\put(0,0){\includegraphics[width=\unitlength,page=1]{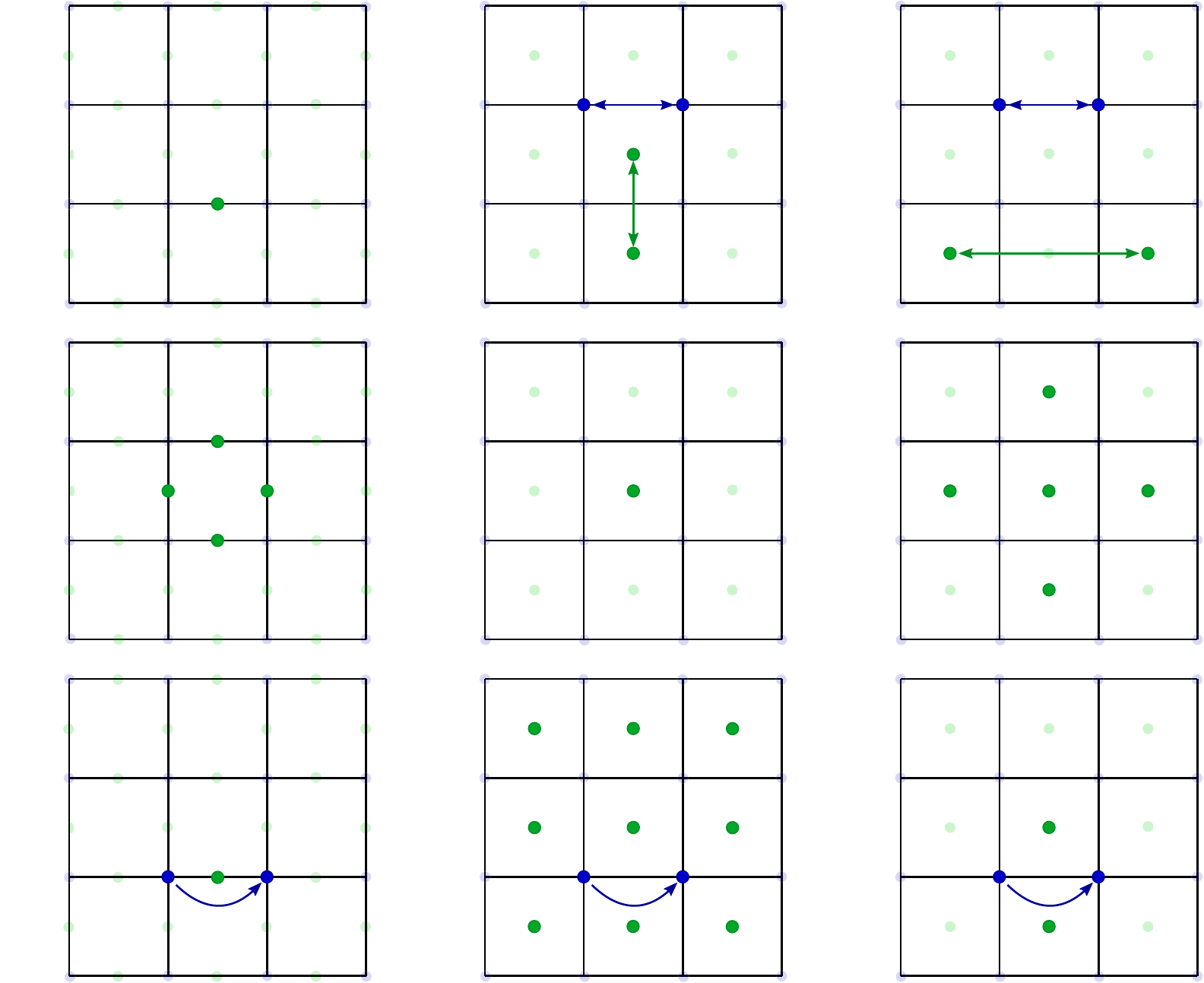}}%
	\put(0.17374823,0.09809543){\makebox(0,0)[lt]{\lineheight{1.25}\smash{\begin{tabular}[t]{l}$+1$\end{tabular}}}}%
	\put(0.10397039,0.07015214){\makebox(0,0)[lt]{\lineheight{1.25}\smash{\begin{tabular}[t]{l}$\Psi(\mathbf{x}) \hspace{44pt} \Psi^{\dagger}(\mathbf{x}+\mathbf{e}_i) \hspace{79pt} \Psi(\mathbf{x}) \hspace{44pt} \Psi^{\dagger}(\mathbf{x}+\mathbf{e}_i) \hspace{78pt} \Psi(\mathbf{x}) \hspace{44pt} \Psi^{\dagger}(\mathbf{x}+\mathbf{e}_i)$\\\\\end{tabular}}}}%
	\put(-0.18097778,0.85264209){\makebox(0,0)[lt]{\lineheight{1.25}\smash{\begin{tabular}[t]{l}-\end{tabular}}}}%
	\put(0.44919412,0.04029772){\makebox(0,0)[lt]{\lineheight{1.25}\smash{\begin{tabular}[t]{l}$-\frac{1}{28}$\end{tabular}}}}%
	\put(0.6136837,0.04021278){\makebox(0,0)[lt]{\lineheight{1.25}\smash{\begin{tabular}[t]{l}$-\frac{1}{28}$\end{tabular}}}}%
	\put(0.53126468,0.04037747){\makebox(0,0)[lt]{\lineheight{1.25}\smash{\begin{tabular}[t]{l}$-\frac{23}{112}$\end{tabular}}}}%
	\put(0.6138557,0.12243567){\makebox(0,0)[lt]{\lineheight{1.25}\smash{\begin{tabular}[t]{l}$+\frac{1}{16}$\end{tabular}}}}%
	\put(0.53143672,0.12260031){\makebox(0,0)[lt]{\lineheight{1.25}\smash{\begin{tabular}[t]{l}$+\frac{1}{4}$\end{tabular}}}}%
	\put(0.61381791,0.20467617){\makebox(0,0)[lt]{\lineheight{1.25}\smash{\begin{tabular}[t]{l}$+\frac{1}{28}$\end{tabular}}}}%
	\put(0.44932836,0.20476111){\makebox(0,0)[lt]{\lineheight{1.25}\smash{\begin{tabular}[t]{l}$+\frac{1}{28}$\end{tabular}}}}%
	\put(0.4493662,0.12252066){\makebox(0,0)[lt]{\lineheight{1.25}\smash{\begin{tabular}[t]{l}$+\frac{1}{16}$\end{tabular}}}}%
	\put(0.53139897,0.20484086){\makebox(0,0)[lt]{\lineheight{1.25}\smash{\begin{tabular}[t]{l}$+\frac{9}{112}$\end{tabular}}}}%
	\put(0.86451011,0.10661056){\makebox(0,0)[lt]{\lineheight{1.25}\smash{\begin{tabular}[t]{l}$+1$\end{tabular}}}}%
	\put(0.8645621,0.02431278){\makebox(0,0)[lt]{\lineheight{1.25}\smash{\begin{tabular}[t]{l}$-1$\end{tabular}}}}%
	\put(0.53148319,0.40200145){\makebox(0,0)[lt]{\lineheight{1.25}\smash{\begin{tabular}[t]{l}$\hspace{1pt}+1$\end{tabular}}}}%
	\put(0.95921525,0.40210164){\makebox(0,0)[lt]{\lineheight{1.25}\smash{\begin{tabular}[t]{l}$ \hspace{1pt} -1$\end{tabular}}}}%
	\put(0.87679193,0.40203038){\makebox(0,0)[lt]{\lineheight{1.25}\smash{\begin{tabular}[t]{l}$\hspace{1pt}+4$\end{tabular}}}}%
	\put(0.86455052,0.50023855){\makebox(0,0)[lt]{\lineheight{1.25}\smash{\begin{tabular}[t]{l}$-1$\end{tabular}}}}%
	\put(0.75371351,0.40227454){\makebox(0,0)[lt]{\lineheight{1.25}\smash{\begin{tabular}[t]{l}$\hspace{4pt} -1$\end{tabular}}}}%
	\put(0.86455479,0.30401224){\makebox(0,0)[lt]{\lineheight{1.25}\smash{\begin{tabular}[t]{l}$-1$\end{tabular}}}}%
	\put(0.17370179,0.45905826){\makebox(0,0)[lt]{\lineheight{1.25}\smash{\begin{tabular}[t]{l}$-1$\end{tabular}}}}%
	\put(0.22693507,0.40205345){\makebox(0,0)[lt]{\lineheight{1.25}\smash{\begin{tabular}[t]{l}$\hspace{1pt}+1$\end{tabular}}}}%
	\put(0.1736399,0.34507273){\makebox(0,0)[lt]{\lineheight{1.25}\smash{\begin{tabular}[t]{l}$+1$\end{tabular}}}}%
	\put(0.10401526,0.40233459){\makebox(0,0)[lt]{\lineheight{1.25}\smash{\begin{tabular}[t]{l}$\hspace{4pt} -1$\end{tabular}}}}%
	\put(0.17485814,0.65695817){\makebox(0,0)[lt]{\lineheight{1.25}\smash{\begin{tabular}[t]{l}$E_i(\mathbf{x})$\end{tabular}}}}%
	\put(0.44379975,0.71215631){\makebox(0,0)[lt]{\lineheight{1.25}\smash{\begin{tabular}[t]{l}$Q(\mathbf{x'}) \hspace{44pt} Q(\mathbf{y'}) \hspace{93pt} Q(\mathbf{x'}) \hspace{44pt} Q(\mathbf{y'})$\end{tabular}}}}%
	\put(-0.00054989,0.68142417){\makebox(0,0)[lt]{\lineheight{1.25}\smash{\begin{tabular}[t]{l}$\mathbf{H_E}$\end{tabular}}}}%
	\put(-0.00053836,0.40214746){\makebox(0,0)[lt]{\lineheight{1.25}\smash{\begin{tabular}[t]{l}$\mathbf{H_B}$\end{tabular}}}}%
	\put(-0.00041768,0.12218076){\makebox(0,0)[lt]{\lineheight{1.25}\smash{\begin{tabular}[t]{l}$\mathbf{H_{int}}$\end{tabular}}}}%
	\put(0.47924363,0.73995323){\makebox(0,0)[lt]{\lineheight{1.25}\smash{\begin{tabular}[t]{l}$\hspace{3pt} V_{\mathrm{Coul}}(\mathbf{x'},\mathbf{y'}) \hspace{134.5pt} V_{\mathrm{Coul}}(\mathbf{x'},\mathbf{y'})$\end{tabular}}}}%
	\put(0.53130634,0.68189963){\makebox(0,0)[lt]{\lineheight{1.25}\smash{\begin{tabular}[t]{l}$ \hspace{1pt} L(\mathbf{x})$\end{tabular}}}}%
	\put(0.47211227,0.58355948){\makebox(0,0)[lt]{\lineheight{1.25}\smash{\begin{tabular}[t]{l}$\hspace{11pt} L(\mathbf{x}-\mathbf{e}_j)$\end{tabular}}}}%
	\put(0.7834066,0.58350834){\makebox(0,0)[lt]{\lineheight{1.25}\smash{\begin{tabular}[t]{l}$ M(\mathbf{x})$\end{tabular}}}}%
	\put(0.94825924,0.58352158){\makebox(0,0)[lt]{\lineheight{1.25}\smash{\begin{tabular}[t]{l}$ M(\mathbf{y})$\end{tabular}}}}%
	\put(-0.18097778,0.85264209){\makebox(0,0)[lt]{\lineheight{1.25}\smash{\begin{tabular}[t]{l}$$\end{tabular}}}}%
	\put(0.83954007,0.58456807){\makebox(0,0)[lt]{\lineheight{1.25}\smash{\begin{tabular}[t]{l}$V_\mathrm{Coul}(\mathbf{x},\mathbf{y})$\end{tabular}}}}%
	\end{picture}%
	\endgroup%
	\caption{Illustration of the interactions in the Hamiltonian for a $3 \times 3$ lattice with open boundary conditions in the original formulation (left column), the dual formulation in terms of $B/L$-variables (middle column) and the dual formulation in terms of $\theta/M$-variables (right column). We consider the electric Hamiltonian (upper row), the magnetic Hamiltonian (middle row) and the gauge-matter interactions (lower row). We leave out the mass part since it is not altered by the transformation so that the only difference is that the matter degrees of freedom participating in it are not subject to local constraints anymore. While the electric Hamiltonian in the original formulation is the sum over the square of the electric field on every link, in the dual picture it is split into a longitudinal part and a transverse part. The longitudinal part of the electric field is completely fixed by the charge configuration which gives in the rotated frame rise to Coulomb interactions (given by the Green's function $G(\mathbf{x},\mathbf{y})$ but named $V_{\mathrm{Coul}}(\mathbf{x},\mathbf{y})$ in the figure for illustrative purposes) between charges $Q(\mathbf{x})$ and $Q(\mathbf{y})$. The transverse part depends on the dual formulation: For the $B/L$-variables the transverse electric field is just the lattice curl of $L$ so that the electric Hamiltonian on a link involves only the two neighboring plaquettes. For the $\theta/M$-variables one can show that the transverse electric Hamiltonian leads to Coulomb interactions among the $M$-variables, thus generating very similar interactions as between the charges. The magnetic Hamiltonian in the original formulation is a four-body interaction among the links: it is the sum of a raising and a lowering operator of the electric field around the plaquette (in the figure we show the effect of the raising operator in the electric basis). In terms of the $B/L$-variables, since $B$ is the lattice curl of the gauge field around a plaquette, it is a one-body term, raising the $L$-variable by one. In terms of the $\theta/M$-variables, it is a five-body interaction, corresponding to the (negative) Laplacian of $\theta$, which raises the $M$-variable in the center by four and lowers it on the neighboring plaquettes by one. The gauge-matter interactions describe the effect of a hopping matter degree of freedom on the gauge field. In the original formulation, the electric field along the link gets raised by one (lowered for hopping in the other direction). In the rotated frame, only changes in the transverse part of the electric field need to be taken into account. In terms of the $B/L$-variables, since the $L$-variables generate the transverse part of the electric field, this change can be expressed by shifting the $L$-variables by the proper amount. These shifts $s_{\mathbf{x},i}(\mathbf{y})$ are shown in the figure (for their calculation see Appendix \ref{helmholtz}). They decay away from the link where hopping occurs, which can already be seen on the $3 \times 3$ lattice. In terms of the $\theta/M$-variables, since the curl of the electric field is only affected on the two neighboring plaquettes, the gauge-matter interactions become local.}
	\label{interactions_figure}
\end{figure*}
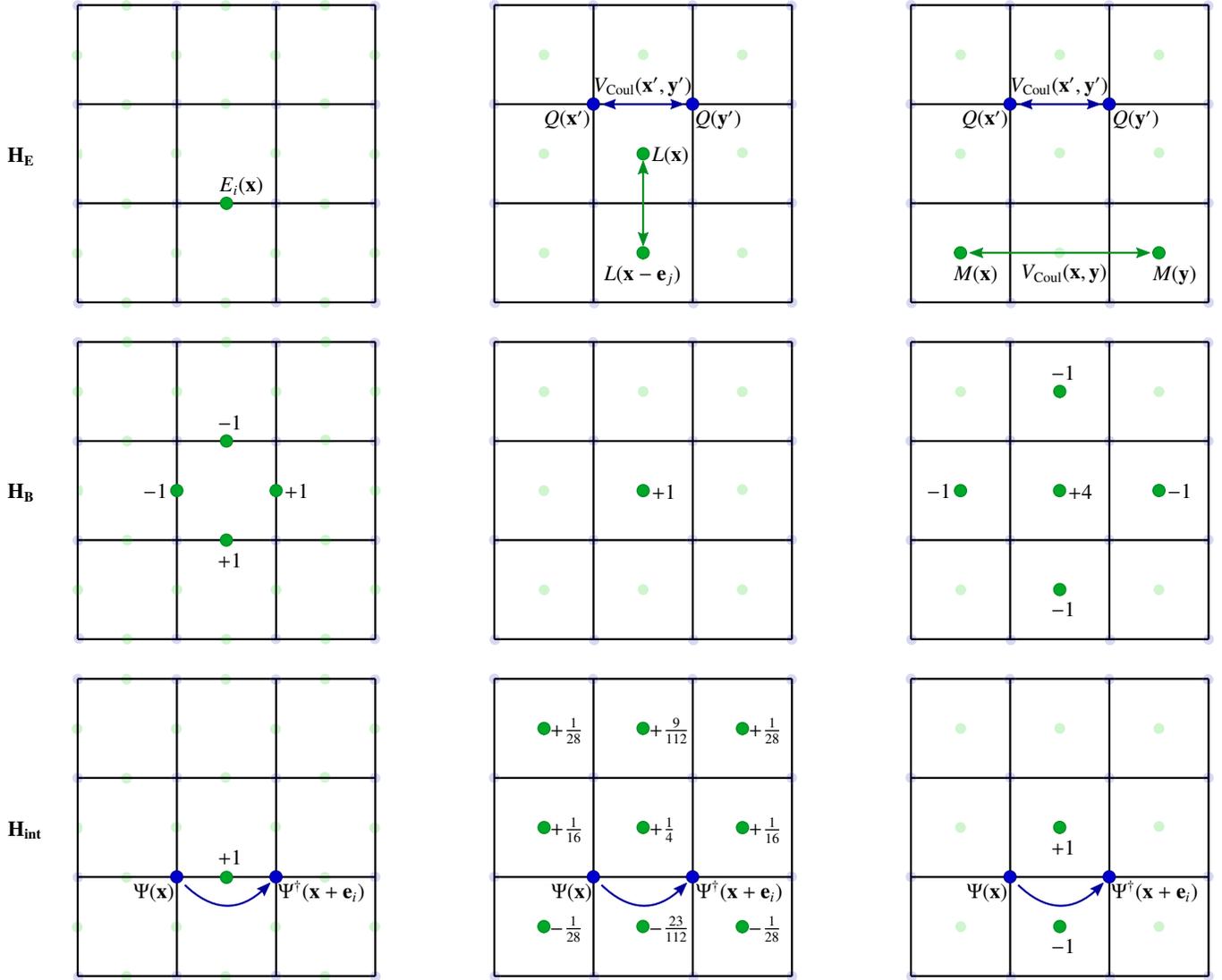

To express the transverse part of the electric energy in terms of the $M$-variables,
we need to find a relation between the $M$- and $L$-variables which can then be inserted in the formula for $\tilde{H}_E^T$ in terms of $L$, eq.~(\ref{etransL}). Such a relation can be obtained by plugging the electric field in terms of $L$, eq.~(\ref{EL}), into the definition of $M$, eq.~(\ref{ME}):
\begin{equation}
M(\mathbf{x})= -\nabla^2 L (\mathbf{x}) + \epsilon_{ij} \Delta_{i}^{(+)} \delta_{x_i,0} L_j
\label{MintermsofL}
\end{equation}   
The second term is a boundary term coming from the global loops (only present for periodic boundary conditions). For open boundary conditions, the relation contains only the first term, leading to a Poisson equation on the plaquettes. Thus, for open boundary conditions, $L$ can be expressed in terms of $M$ by the Green's function. Inserting this in eq.~(\ref{etransL}) and using the lattice analog of integrating by parts, gives a Coulomb interaction between the $M$-variables. For periodic boundary conditions, this interaction potential is slightly modified by boundary effects due to the second term in eq.~(\ref{MintermsofL}), i.e.
\begin{equation} \label{extraCoulomb}
\tilde H^T_E = \frac{g^2}{2}\underset{\mathbf{x},\mathbf{y}}{\sum}
M\left(\mathbf{x}\right)
 \widetilde{G}\left(\mathbf{x},\mathbf{y}\right)  
M\left(\mathbf{y}\right)
\end{equation}
where $\widetilde{G}\left(\mathbf{x},\mathbf{y}\right)$ denotes the modified interaction potential. Note that it also includes interactions with the two global variables, i.e. the sum above contains also $M_1$ and $M_2$ (for its exact form see Appendix \ref{details}). For open boundary conditions,
$\widetilde{G}\left(\mathbf{x},\mathbf{y}\right)$ reduces to $G\left(\mathbf{x},\mathbf{y}\right)$.
Thus, the interactions between the curls of the electric field on the plaquettes are of Coulomb type; it shows some strong similarity with the longitudinal part $ \tilde H^L_E$ in eq.~(\ref{HL}) where exactly the same interaction appears between the matter degrees of freedom. The last remaining part is the magnetic Hamiltonian. If we express $B$ in terms of $\theta$ in the same way as we did for $M$ in terms of $L$ in eq.~(\ref{MintermsofL}), we can write down $\tilde{H}_B$ in terms of $\theta$, following eq.~(\ref{magnetichamiltonianintermsofB}): 
\begin{align}
\tilde H_B =& -\frac{1}{g^2}\underset{\mathbf{x}}{\sum}\cos\left(-\nabla^2\theta\left(\mathbf{x}\right) + \epsilon_{ij} \Delta_{i}^{(+)} \delta_{x_i,0} \theta_j \right) \\
=& -\frac{1}{g^2}\underset{\mathbf{x}}{\sum}\cos\Big(4\theta\left(\mathbf{x}\right)-\theta\left(\mathbf{x}+\hat{\mathbf{e}}_1\right)
-\theta\left(\mathbf{x}+\hat{\mathbf{e}}_2\right) \nonumber\\
&\phantom{-\frac{1}{g^2}\underset{\mathbf{x}}{\sum}\cos\big(} 
-\theta\left(\mathbf{x}-\hat{\mathbf{e}}_1\right) 
-\theta\left(\mathbf{x}-\hat{\mathbf{e}}_2\right) \nonumber\\
&\phantom{-\frac{1}{g^2}\underset{\mathbf{x}}{\sum}\cos\big(} 
+ \left( \delta_{x_2,0} - \delta_{x_2,-1} \right) \theta_1
- \left( \delta_{x_1,0}
- \delta_{x_1,-1} \right)  \theta_2
\Big) \nonumber
\end{align}
where the contributions of the global variables $\theta_1$ and $\theta_2$ are only present on plaquettes sharing a link with the $\mathbf{\hat{e}}_1$- or $\mathbf{\hat{e}}_2$-axis. They vanish completely for open boundary conditions and the magnetic Hamiltonian becomes a local interaction - a five-body one, involving a plaquette and its neighbors. The action of $\tilde{H}_B$ in the electric basis is illustrated in Fig.~\ref{interactions_figure}, for both the original and the dual formulations.

Analogous to the dual formulation with $B$ and $L$, there is also a global constraint for the formulation in terms of $\theta$ and $M$. This constraint on physical states can be obtained if we sum eq.~(\ref{ME}) over the whole lattice (again, this only holds true for periodic boundary conditions, it is not the case for open boundary conditions),
\begin{equation}
\underset{\mathbf{x}}{\sum}M\left(\mathbf{x}\right) \left|\mathrm{phys}\right\rangle= 0
\label{Mglob}
\end{equation}

Overall, the original formulation of the lattice gauge theory has been replaced in this dual formulation of the transformed picture by (assuming periodic boundary conditions)
\begin{widetext}
\begin{equation}
\begin{aligned}
\tilde{H} =& H_m 
 +\frac{g^2}{2}\underset{\mathbf{x},\mathbf{y}}{\sum}\left[
\left(Q\left(\mathbf{x}\right)+q\left(\mathbf{x}\right)\right)
G\left(\mathbf{x},\mathbf{y}\right)
\left(Q\left(\mathbf{y}\right)+q\left(\mathbf{y}\right)\right) +
M\left(\mathbf{x}\right)
 \widetilde{G}\left(\mathbf{x},\mathbf{y}\right) 
M\left(\mathbf{y}\right)
\right]
\\&-\frac{1}{g^2}\underset{\mathbf{x}}{\sum}\cos\left(-\nabla^2\theta\left(\mathbf{x}\right) + \epsilon_{ij} \Delta_{i}^{(+)} \delta_{x_i,0} \theta_j \right)
+ \underset{\mathbf{x},i}{\sum} t_{\mathbf{x},i}\Psi^{\dagger}\left(\mathbf{x}\right)e^{i \left( \epsilon_{ij}\Delta^{(-)}_j\theta\left(\mathbf{x} \right)+\delta_{x_j,0} \theta_i \right)}
\Psi\left(\mathbf{x}+\hat{\mathbf{e}}_i\right)+h.c.
\end{aligned}
\end{equation}
\end{widetext}
with $i \neq j$. The link variables, the angle $\phi_i\left(\mathbf{x}\right)$ and the integer-valued $E_i\left(\mathbf{x}\right)$, and the multiple local constraints imposed by the Gauss law (\ref{Gauss}) are replaced by the dual plaquette variables, the angle $\theta\left(\mathbf{x}\right)$ and the integer-valued $M\left(\mathbf{x}\right)$, and the single global constraint (\ref{Mglob}). For open boundary conditions, the formulation simplifies since the modified interaction potential $\widetilde{G}\left(\mathbf{x},\mathbf{y}\right) $ reduces to $G\left(\mathbf{x},\mathbf{y}\right)$ and the global loop contributions corresponding to $\theta_1$ and $\theta_2$ drop out of the gauge-matter interactions and the magnetic interactions, rendering them completely local. Moreover, there is no global constraint left, making the formulation manifestly gauge-invariant. The reason why periodic boundary conditions are more difficult to deal with in the presence of dynamical matter as compared to static matter is that the two global loops around the torus become dynamical due to the appearance of gauge-matter interactions. Thus, the choice of open boundary conditions should be preferred, in particular for quantum simulations as open boundary conditions are much more natural from an experimental point of view. To summarize, the required interactions for open boundary conditions in the original approach, the formulation in terms of $B$/$L$-variables and in terms of $\theta$/$M$-variables are illustrated in Fig.~\ref{interactions_figure}, exemplary for a $3 \times 3$ lattice.

\section{Three Space Dimensions} \label{threedimensions}
In this section we consider the generalization of the previous discussion to $3+1d$, i.e. three space dimensions. Difference operators are defined exactly in the same manner as in the two dimensional settings of section \ref{latsec}, as well as the gradient and the divergence. The Laplacian's definition changes by a numerical factor, to
\begin{equation}
\nabla^2 f\left(\mathbf{x}\right) = \Delta^{(-)}_i\Delta^{(+)}_i f\left(\mathbf{x}\right) 
= \underset{i=1,2,3}{\sum}\left(f\left(\mathbf{x}+\hat{\mathbf{e}}_i\right)-f\left(\mathbf{x}-\hat{\mathbf{e}}_i\right)\right) - 6f\left(\mathbf{x}\right)
\end{equation}
We need to generalize the definitions of the curl. The curl of a vector field on the links will be a pseudovector field residing at the centers of plaquettes,
\begin{equation}
\left(\nabla\times\mathbf{F}\left(\mathbf{x}\right)\right)_i=\epsilon_{ijk}\Delta^{(+)}_jF_k\left(\mathbf{x}\right)
\end{equation}
while the curl of a pseudovector will be a regular vector field on the links,
\begin{equation}
\left(\nabla\times\mathbf{L}\left(\mathbf{x}\right)\right)_i=\epsilon_{ijk}\Delta^{(-)}_jL_k\left(\mathbf{x}\right)
\end{equation}
If we fix $i=3$, we recover the expressions for two space dimensions.

The original Hamiltonian $H$ in eq.~(\ref{originalhamiltonian}) is straightforwardly generalized: $H_m$ still runs over all lattice sites, $H_E$ and $H_{int}$ run over all links (each vector now contains three components) and $H_B$ now includes three differently oriented plaquette interactions (not a single one as for two space dimensions), taking the form
\begin{equation}
H_B = -\frac{1}{g^2}\underset{\mathbf{x},i}{\sum}\cos\left(\epsilon_{ijk}\Delta_j^{\left(+\right)}\phi_k\left(\mathbf{x}\right)\right)
\label{HamiltonianB}
\end{equation}
which we will express in terms of the magnetic field variables
\begin{equation}
B_i\left(\mathbf{x}\right)=\epsilon_{ijk}\Delta_j^{\left(+\right)}\phi_k\left(\mathbf{x}\right)
\label{Magneticfield3d}
\end{equation}
The Gauss law takes the same form (\ref{Gauss}), this time with the three-dimensional divergence.

Decoupling the matter can be done with the same transformation $\mathcal{U}$ used in section \ref{decoupsec}, now in a three dimensional space and with the $d=3$ Green's function (see Appendix \ref{greensfunction}) instead of the two dimensional one used above. Most of the transformed parts of the Hamiltonian ($\tilde H_E^{T}$, $\tilde H_E^{L}$ and $\tilde H_{int}$) will have an identical form as in the $d=2$ case with the straightforward dimensional generalization, see eqs. (\ref{HT}), (\ref{HL}) and (\ref{Hinttilde}). Also the Gauss law transforms in the same manner, i.e. it gets decoupled from the matter degrees of freedom as in eq. (\ref{GaussT}). $H_m$ and $H_B$ still commute with the transformation $\mathcal{U}$. 

The crucial difference in three space dimensions appears when formulating the transformed Hamiltonian in terms of dual variables. We start with the dual formulation in terms of the $B$- and $L$-variables in section \ref{dualformulation}. From eq.~(\ref{Magneticfield3d}) it is clear that the divergence of $B$ is zero, i.e. $\Delta_{i}^{(+)} B_{i}(\mathbf{x})= 0  \hspace{2pt} \forall \hspace{1pt} \mathbf{x}$ holds on the operator level. Since this is an operator identity it is satisfied by any state; however, when building a classical or quantum simulation it cannot be assumed to be satisfied a priori and thus physical states need to fulfill a constraint for every cube:

\begin{equation}
\Delta_{i}^{(+)} B_{i}(\mathbf{x}) \left|\mathrm{phys}\right\rangle=0  \quad \forall \hspace{1pt} \mathbf{x}
\label{localconstraint3dB}
\end{equation} 
One can intuitively think about it in the electric basis, as raising the electric flux on all faces of a cube should leave the state invariant. Therefore, in three dimensions there are local constraints left. However, they do not involve the matter degrees of freedom. Note that for periodic boundary conditions these local constraints are not independent, since the sum over all local constraint gives zero, i.e. there are $N^3-1$ independent constraints. This is not the case for open boundary conditions. In addition there are three global constraints, which are a generalization of the single global constraint in two dimensions (again, only for periodic boundary conditions):

\begin{equation}
\sum_{\substack{x_i=0 \\ x_j,x_k}} B_i(\mathbf{x}) \left|\mathrm{phys}\right\rangle = 0  \quad \mathrm{for} \quad i=1,2,3
\label{globalconstraint3dBL}
\end{equation}
with $i\neq j \neq k$ and $\mathbf{x}=(x_1,x_2,x_3)$. These global constraints correspond to slices through the lattice. There are only three independent ones since all other slices can be obtained by deforming them with the plaquette constraints from eq.~(\ref{localconstraint3dB}).  

We express the (transverse) electric field $E_i\left(\mathbf{x}\right)$ after the transformation as the curl of a pseudovector $L_k\left(\mathbf{x}\right)$. Thus, the definition of eq. (\ref{EL}) is replaced by
\begin{equation}
E_i\left(\mathbf{x}\right)=\epsilon_{ijk}\Delta^{(-)}_jL_k\left(\mathbf{x}\right) + \delta_{x_j,0} \delta_{x_k,0} L_i.
\label{EL3}
\end{equation}
with $i \neq j \neq k$. The $B_1/L_1$-, $B_2/L_2$- and $B_3/L_3$-variables correspond to the three global loops winding around the lattice along a specific axis (only present for periodic boundary conditions). A discussion of the arising topological phenomena due to these global loops can be found in \cite{meurice2020abeliangauge}. 

One can show that, similar to the two-dimensional case, the $B$- and $L$-variables fulfill canonical commutation relations,
\begin{equation}
\left[B_i\left(\mathbf{x}\right),L_j\left(\mathbf{y}\right)\right]=i\delta_{ij}\delta(\mathbf{x},\mathbf{y}).
\end{equation}

Similar to the two dimensional case, we can perform a counting of degrees of freedom. For that we can neglect the matter degrees of freedom since their number is the same in both formulations. In the case of periodic boundary conditions, we have in the original link formulation $3N^3$ links and $N^3$ sites, thus $N^3-1$ independent Gauss laws, leading to $2N^3+1$ physical gauge degrees of freedom. In the dual formulation, there are $3N^3$ plaquettes, three global loop variables winding around the lattice,  $N^3-1$ independent cube constraints as in eq.~(\ref{localconstraint3dB}) and three global constraints, giving also a total of $2N^3+1$ physical gauge degrees of freedom.

In the case of open boundary conditions, we have in the original formulation $3N(N+1)^2$ links and $(N+1)^3$ sites, i.e $(N+1)^3-1$ Gauss law constraints and thus $2N^3+3N^2$ physical gauge degrees of freedom. In the dual formulation, we have $3N^2(N+1)$ plaquettes and $N^3$ cube constraints, resulting also in $2N^3+3N^2$ physical gauge degrees of freedom.

The (transverse) electric Hamiltonian written in terms of the $L$-variables takes a similar form as in the two-dimensional case,

\begin{equation}
\tilde{H}_E^T=\frac{g^2}{2}\sum_{\mathbf{x},i} \left(\epsilon_{ijk} \Delta_j^{(-)} L_k(\mathbf{x}) + \delta_{x_j,0} \delta_{x_k,0} L_i \right)^2
\end{equation}
with $i \neq j \neq k$ (the second term vanishes for open boundary conditions). The difference in three dimensions is that $L$-variables on four plaquettes (the ones containing the link) are required to express the transverse part of the electric field. The magnetic Hamiltonian is still a one-body term, as in two dimensions, which can be seen from eq. (\ref{HamiltonianB}) and (\ref{Magneticfield3d}). The gauge-matter interactions have the same form as in (\ref{dualgaugematter2d}) with the difference that the shifts $s_{\mathbf{x},i}(\mathbf{y})$ in the $L(\mathbf{y})$ variables in (\ref{expressphiTwithshifts}) are computed with the three-dimensional Green's function (see Appendix \ref{greensfunction}). Although this interaction involves many degrees of freedom, the shifts decay away from the link ($\mathbf{x},i$) (and even faster in three dimensions) which might allow one to neglect contributions above some certain distance. 

The dual formulation in terms of the $\theta$- and $M$-variables can be generalized in a similar fashion. We first define a pseudovector field on the plaquettes, $\theta_k\left(\mathbf{x}\right)$, whose curl generates the transverse part of the gauge field (in addition to the global loop variables $\theta_i$):
\begin{equation}
\phi^T_i\left(\mathbf{x}\right)=\epsilon_{ijk}\Delta^{(-)}_j\theta_k\left(\mathbf{x}\right) +  \delta_{x_j,0} \delta_{x_k,0} \theta_i.
\end{equation}
with $i \neq j \neq k$ (the second term vanishes for open boundary conditions). We also define $M$-variables as the curl of the electric field

\begin{equation}
M_i\left(\mathbf{x}\right)=\epsilon_{ijk}\Delta^{(+)}_j E_k\left(\mathbf{x}\right)
\end{equation}
With the same reasoning as for the $B$-variables, we obtain similar local constraints as in eq.~(\ref{localconstraint3dB}) for the $M$-variables

\begin{equation}
\Delta_{i}^{(+)} M_{i}(\mathbf{x}) \left|\mathrm{phys}\right\rangle=0  \quad \forall \hspace{1pt} \mathbf{x}
\end{equation} 
and the commutation relations 

\begin{equation}
\left[\theta_i\left(\mathbf{x}\right),M_j\left(\mathbf{y}\right)\right]=i\delta_{ij}\delta(\mathbf{x},\mathbf{y}).
\end{equation}
As in the two dimensional case, the operators $M_i\left(\mathbf{x}\right)$ have an integer spectrum. For periodic boundary conditions, the physical states need to fulfill the global constraints in eq.(\ref{globalconstraint3dBL}), with $B$ replaced by $M$. The counting of degrees of freedom can be performed in the same way as for the $B$/$L$-variables.

The gauge-matter interactions written in terms of the $\theta$-variables result again in local interactions (up to contributions from the global loop variables $\theta_i$ which are only present for periodic boundary conditions and then only on the axes),

\begin{equation}
\tilde{H}_{int}=\underset{\mathbf{x},i}{\sum} t_{\mathbf{x},i} \Psi^{\dagger}\left(\mathbf{x}\right)e^{i \left( \epsilon_{ijk} \Delta^{(-)}_j\theta_k\left(\mathbf{x}\right) + \delta_{x_j,0} \delta_{x_k,0} \theta_i \right)}
\Psi\left(\mathbf{x}+\hat{\mathbf{e}}_i\right)+h.c.
\end{equation}
 with $i \neq j \neq k$. In three dimensions four plaquettes are contributing compared to two in the two-dimensional case. If we express the magnetic interactions in terms of the $\theta$-variables, we obtain

\begin{equation}
H_{B}=-\frac{1}{g^2}\sum_{\mathbf{x},i} \cos\left( \epsilon_{ijk} \epsilon_{klm} \Delta^{(+)}_j \Delta^{(-)}_l \theta_m(\mathbf{x}) \right).
\end{equation}
The magnetic interaction on a plaquette involves all $\theta$-variables which share a link with the respective plaquette. 

To conclude, in three space dimensions, the matter degrees of freedom can be decoupled from the gauge constraints, so that only the gauge field variables on the plaquettes are subject to constraints. However, compared to two dimensions, the remaining constraints are local, i.e. every cube on the lattice defines such a constraint. It involves six plaquette variables, compared to six link variables and the charge on the site in the original Gauss law. On the other hand, due to the additional dimension more degrees of freedom participate in the interactions, making the dual formulation in three dimensions more difficult to study compared to the two-dimensional version.

\section{Discussion and Conclusions}

In this work, we have shown how to unitarily transform compact QED with dynamical matter to a frame, in which physical states (i.e. states fulfilling the local gauge constraints) can be expressed by dual, gauge-invariant variables while keeping translational invariance. The central concept in this transformation is the decomposition of lattice vector fields into transverse and longitudinal components (Helmholtz decomposition). In the original formulation, the gauge constraints (Gauss laws) for physical states fix the longitudinal component of the electric field by the given charge configuration. The transverse component is not affected by these constraints. Since the gauge-matter interactions (which appear due to the presence of dynamical matter) involve the longitudinal part of the gauge field (the canonically conjugate variable to the electric field) we transform to a rotated frame, where Coulomb gauge holds, i.e. only the transverse component of the gauge field appears in the Hamiltonian. In this transformed picture, the matter degrees of freedom decouple from the Gauss laws and the physical (transverse) part of the gauge field and the electric field is expressed in terms of a new set of canonical variables on the plaquettes, making the formulation manifestly gauge-invariant.
 
The transformation can be performed in two and three spatial dimensions, with periodic and open boundary conditions and there are two sets of dual variables, in terms of which one can express the transverse part of the gauge field/electric field.

While the unitary transformation in two and three spatial dimensions has a very similar form, the formulation in terms of dual variables is quite different. In two dimensions, the dual plaquette variables are completely free of any local constraints. In three dimensions, however, there exists a local constraint for every cube on the lattice, in which all plaquette variables on it are involved. This is related to the fact that every closed surface defines a constraint for these dual variables because a transverse field (a curl field) integrated over a closed surface needs to be zero. Nevertheless, these local constraints only involve the gauge field and not the matter degrees of freedom. 

The main difference in boundary conditions is that in the case of periodic boundary conditions there are global loop variables around the lattice for every spatial direction, which are not present for open boundary conditions. The introduction of \emph{dynamical} charges and therefore the appearance of gauge-matter interactions makes these variables dynamical which is a big difference compared to the case of static matter where these variables would just fix a topological sector \cite{kaplan_gauss_2018}. Also, for periodic boundary conditions there are additional closed surfaces (in $2+1d$ the whole torus), giving rise to global constraints on the dual plaquette variables.   

The two sets of dual variables share the same locations for their degrees of freedom, the difference between them arises in the complexity of the different terms in the Hamiltonian. The terms where no gauge field is involved are the same, namely the mass term for the matter and the (longitudinal) electric Hamiltonian, which is after the transformation a Coulomb interaction of (static and dynamical) charges. While the dual formulation in terms of $B$/$L$-variables (see section \ref{dualformulation}) makes the magnetic Hamiltonian a one-body term and the (transverse) electric Hamiltonian a local two-body interaction, the gauge-matter interactions are more complicated since the hopping of a matter degree of freedom affects many plaquette variables on the lattice. On the other hand, the dual formulation in terms of $\theta$/$M$-variables makes the gauge-matter interactions local again, only involving the plaquette variables containing the link where hopping occurs. The magnetic interaction becomes (focusing on two dimensions now) a five-body interaction among a plaquette and its four neighbors. The (transverse) electric Hamiltonian becomes a Coulomb interaction among the plaquette variables. 

Both formulations could be useful for both classical variational studies or quantum simulation and computation of lattice gauge theories, where descriptions with no or less gauge redundancies help to reduce the required resources and prevent a possible violation of gauge-invariance. For variational studies, this also allows one to consider a larger class of possible ansatz states (due to the absence of constraints). 

The $B$/$L$-formulation could be used to extend variational ansatz states for compact QED with static matter (as in \cite{PhysRevResearch.2.043145}) to dynamical charges, e.g. by coupling it to a fermionic Gaussian state with a Non-Gaussian transformation \cite{shi2018variational}. In this context, a recent Monte Carlo study \cite{prx2020assad} could serve as a useful benchmark, as it provides results for an even number of fermion flavors where the sign problem is absent.

The $\theta$/$M$-formulation could be used to design a quantum simulation, where the difficult terms in the implementation are Coulomb interactions and the five-body interaction corresponding to the magnetic Hamiltonian. However, the latter is nothing but the ordinary four-body (plaquette) interaction in the Kogut-Susskind Hamiltonian with an additional degree of freedom in the center of the respective plaquette. Over the last years, there has been a lot of effort in how to implement this interaction in a quantum simulation \cite{tagliacozzo_optical_2013,tagliacozzo_simulation_2012,zohar_digital_2017,zohar_digital_2017-1, bender_digital_2018} which can be a nice starting point for the implementation of the interaction above. The more difficult part is generating a Coulomb potential with quantum devices. This problem is shared with quantum simulation of quantum chemistry where a Coulomb potential is a crucial building block. However, recently, there have been ideas how to implement such a potential with ultracold atoms \cite{arguello2019analogue} which might also be beneficial to lattice gauge theory. To further reduce the required resources for a quantum simulation, one could combine our approach with a truncation scheme proposed in \cite{haase_resource_2020}, applied to the Hilbert space of the dual variables.

As for other gauge groups, the method could be straightforwardly extended to $\mathbb{Z}_N$ - as they are all subgroups of $U(1)$. For non-Abelian groups the situation is different;
if one tries to decompose the gauge field/electric field in a similar way to the Abelian case, the equations become non-linear (as opposed to Poisson's equation) and thus one cannot perform such a unitary transformation in the same manner as for Abelian gauge groups. There are other methods (e.g. the maximal tree approach \cite{ligterink2000toward}) but they do not preserve translational invariance.

\acknowledgements
We thank David B. Kaplan for fruitful discussions. This research was supported by the Israel Science Foundation (grant No. 523/20). J.B. acknowledges support by the EU-QUANTERA project QTFLAG (BMBF grant No. 13N14780). J.B. thanks the Hebrew University of Jerusalem for the hospitality during his stay at the Racah Institute of Physics.

\appendix

\section{The lattice Poisson equation} \label{greensfunction}

In this section, we discuss the calculation of the lattice Green's function $G(\mathbf{x},\mathbf{y})$ for both periodic and open boundary conditions, defined by the equation 

\begin{equation} \label{definitiongreens}
-\nabla^2 G\left(\mathbf{x},\mathbf{y}\right) = \delta\left(\mathbf{x},\mathbf{y}\right).
\end{equation}

The solutions to Poisson's equation  
\begin{equation}
-\nabla^2 f\left(\mathbf{x}\right) = Q\left(\mathbf{x}\right)
\end{equation}
can be constructed out of it as a superposition,

\begin{equation}
f(\mathbf{x})=\sum_{\mathbf{y}} G(\mathbf{x},\mathbf{y}) Q(\mathbf{y}).
\end{equation}

Starting with periodic boundary conditions in $d$ space dimensions, the (negative) Laplacian takes the form
	\begin{equation} \label{laplacebulk}
- \nabla^2 f\left(\mathbf{x}\right) = 2df\left(\mathbf{x}\right) -  \sum_{i=1}^d\left(f\left(\mathbf{x}+\hat{\mathbf{e}}_i\right)+f\left(\mathbf{x}-\hat{\mathbf{e}}_i\right)\right).
\end{equation}
We define the Fourier transformation on the lattice as 
\begin{equation}
\mathcal{F}\left[f\left(\mathbf{x}\right)\right] = \tilde{f}\left(\mathbf{k}\right)=\frac{1}{N^{d/2}} \underset{\mathbf{x}}{\sum}e^{i \frac{2\pi}{N}\mathbf{k}\cdot\mathbf{x}}
f\left(\mathbf{x}\right).
\end{equation}
We can now obtain the lattice Green's function by Fourier transformation of eq.~(\ref{definitiongreens}), 

\begin{equation} \label{greensfunctionformulaPBC}
G(\mathbf{x},\mathbf{y})=G(\mathbf{x}-\mathbf{y})= \sum_{\mathbf{k} \neq 0}  \frac{e^{i\frac{2\pi}{N} (\mathbf{x}-\mathbf{y}) \mathbf{k} }}{2\left(d-\sum_i \cos \left(\frac{2\pi}{N} k_i\right)\right)} 
\end{equation}
with  $\mathbf{x}=(x_1,..,x_d)$, $\mathbf{k}=(k_1,..,k_d)$ and $x_i,k_i \in \{0,..,N-1\}$. The $\mathbf{k=0}$ mode can be neglected, since the total charge on the lattice is always zero due to gauge invariance. The Green's function in two and three dimensions differs only by an additional term in the denominator in eq.~(\ref{greensfunctionformulaPBC}) due to the additional dimension.

For open boundary conditions, one cannot obtain such an explicit formula due to boundary effects. The lattice sites on the corners only have half the number of neighboring sites compared to the bulk so that the Laplace operator looks different (\ref{laplacebulk}) (say e.g. the bottom left corner, $\mathbf{x=0}$):

\begin{equation}
-\nabla^2 f(\mathbf{0}) = d f(\mathbf{0}) - \sum_{i=1}^d f(\mathbf{0}+\mathbf{\hat{e}}_i)
\end{equation}
The Laplace operator on the edges is modified in an analogous way. Therefore, the operator cannot be diagonalized by a discrete Fourier transform but needs to be inverted numerically. Since the Laplace matrix is singular one needs to fix a condition, e.g. $\sum_{\mathbf{x}} G(\mathbf{x},\mathbf{y})=0$. By fixing $\mathbf{y}$ for different lattice positions and inverting the Laplace matrix, one can then obtain the Green's function $G(\mathbf{x},\mathbf{y})$.

\section{The lattice Helmholtz decomposition} \label{helmholtz}
With the Green's function from the previous section, we can write down the (Helmholtz) decomposition of a lattice vector field into transverse and longitudinal components, as written in eq.~(\ref{hhd}). For that, we will need the double curl identity

\begin{equation}
\begin{aligned}
\left[\nabla\times\left(\nabla\times \mathbf{F}\right)\right]_i \left(\mathbf{x}\right)
&=\epsilon_{ijk}\Delta_j^{(-)}\epsilon_{klm}\Delta_l^{(+)}F_m\left(\mathbf{x}\right)\\
&=(\delta_{il} \delta_{jm} - \delta_{im} \delta_{jl}) \Delta_j^{(-)} \Delta_l^{(+)}  F_m(\mathbf{x})\\
&=\Delta^{(+)}_i \Delta^{(-)}_j F_j(\mathbf{x}) - \Delta_j^{(-)} \Delta_j^{(+)} F_i(\mathbf{x}) \\
&=\nabla_i\left(\nabla\cdot \mathbf{F} \left(\mathbf{x}\right) \right)  - \nabla^2F_i\left(\mathbf{x}\right).
\end{aligned}
\label{doublecurl}
\end{equation}
One should note that for periodic boundary conditions there is an additional contribution in the double curl coming from global loops around the lattice which need to be taken into account. 

\subsection{Periodic boundary conditions}

We can now derive the Helmholtz decomposition in an analogous way to the continuum version (for periodic boundary conditions): 

\begin{equation}
F_i\left(\mathbf{x}\right)=\underset{\mathbf{y}}{\sum}\delta\left(\mathbf{x},\mathbf{y}\right)F_i\left(\mathbf{y}\right)=- \nabla^2_{\mathbf{x}}\underset{\mathbf{y}}{\sum}G\left(\mathbf{x},\mathbf{y}\right)F_i\left(\mathbf{y}\right)
\label{FF}
\end{equation}
Inserting the double curl identity (\ref{doublecurl}), we get the separation into a longitudinal and a transverse component. The longitudinal component has the form

\begin{equation}
F_i^L(\mathbf{x})= -\Delta^{(+)}_i \phi(\mathbf{x})  
\end{equation}
with the scalar field $\phi$ on the sites
\begin{equation} \label{scalarfieldpbc}
\begin{aligned}
\phi\left(\mathbf{x}\right) &=    \sum_{\mathbf{y}} \Delta_{j,\mathbf{x}}^{(-)}  G\left(\mathbf{x},\mathbf{y}\right)F_j\left(\mathbf{y}\right) \\
&=\sum_{\mathbf{y}} G(\mathbf{x},\mathbf{y}) \Delta_{j,\mathbf{y}}^{(-)} F_j(\mathbf{y})
\end{aligned}
\end{equation} 
The transverse component is a little more complicated since we also need to take into account the contributions from the global loops $L_i$ around the lattice. Without the global part, we obtain for the transverse component

\begin{equation}
F_{\mathrm{plaq},i}^{T}(\mathbf{x})=\epsilon_{ijk} \Delta^{(-)}_{j} L_{\mathrm{plaq},k}(\mathbf{x})
\end{equation}

with the pseudovector field $L$ on the plaquettes 
\begin{equation} \label{Lpotentialpbc}
\begin{aligned}
L_{\mathrm{plaq},k}\left(\mathbf{x}\right) &= \sum_\mathbf{y} \epsilon_{klm} \Delta_{l,\mathbf{x}}^{(+)}  G\left(\mathbf{x},\mathbf{y}\right)F_m\left(\mathbf{y}\right) \\ 
&=\sum_\mathbf{y}   G\left(\mathbf{x},\mathbf{y}\right) \epsilon_{klm} \Delta_{l,\mathbf{y}}^{(+)}  F_m\left(\mathbf{y}\right) .
\end{aligned}
\end{equation}
If we look at the field generated by the scalar field $\phi$ and the pseudovector field $L_{\mathrm{plaq}}$ in Fourier space, it is clear that all momentum modes of $F$ can be obtained apart from the $\mathbf{k}=0$ mode, i.e. a constant field. For that, the global loop $L_i$ is required, which needs to be fixed to

\begin{equation} \label{L2}
L_i=\frac{1}{N}\sum_{\mathbf{x}}F_i(\mathbf{x}).
\end{equation}

This gives the correct $\mathbf{k}=0$ mode but in order to get a constant field this contribution needs to be equally distributed over the lattice. Thus, we define an additional $L_{\mathrm{const}}$-field on the plaquettes, on top of $L_{\mathrm{plaq}}$ (exemplary for $L_1$, the other spatial directions follow analogously): 

\begin{equation} \label{L3}
\begin{aligned}
L_{\mathrm{const},3}(\mathbf{x})&= \frac{\sum_{\mathbf{y}} F_1(\mathbf{y})}{N^2} x_2 \quad \mathrm{if} \quad x_3=0 \\
L_{\mathrm{const},2}(\mathbf{x})&= -\frac{\sum_{\mathbf{y}} F_1(\mathbf{y})}{N^3} x_3
\end{aligned}
\end{equation}
$L_{\mathrm{const},3}$ distributes the field of the global loop $L_1$ in the $\mathbf{\hat{e}}_2$-direction and $L_{\mathrm{const},2}$ from the $\mathbf{\hat{e}}_1,\mathbf{\hat{e}}_2$-plane in the $\mathbf{\hat{e}}_3$-direction over the whole lattice, giving us a constant field in the $\mathbf{\hat{e}}_1$-direction. The total plaquette field of the $L$-variables is then $L \equiv L_{\mathrm{plaq}}+L_{\mathrm{const}}$ and the total transverse component $F_i^T(\mathbf{x}) \equiv  F_{\mathrm{plaq},i}^T(\mathbf{x}) + \frac{\sum_{\mathbf{y}} F_i(\mathbf{y})}{N^3} $. The Helmholtz decomposition of $F$ can thus be written as 

\begin{equation}
\begin{aligned}
F_i(\mathbf{x})&= -\Delta^{(+)}_i \phi(\mathbf{x})  +\epsilon_{ijk} \Delta^{(-)}_{j} (L_{\mathrm{plaq}}+L_{\mathrm{const}})_k(\mathbf{x})+ \delta_{x_j,0} \delta_{x_k,0} L_i. \\
&= -\Delta^{(+)}_i \phi(\mathbf{x})  +\epsilon_{ijk} \Delta^{(-)}_{j} L_k(\mathbf{x})+ \delta_{x_j,0} \delta_{x_k,0} L_i. \\
&=F_i^L(\mathbf{x}) + F_{\mathrm{plaq},i}^T(\mathbf{x}) + \frac{\sum_{\mathbf{y}} F_i(\mathbf{y})}{N^3} \\
&=F_i^L(\mathbf{x}) + F_i^T(\mathbf{x})  \\
\end{aligned}
\end{equation}
with $i \neq j \neq k$.

\subsection{Open boundary conditions}

For open boundary conditions, one can perform a similar decomposition, with the major difference that there is no global loop participating. It can be written as 

\begin{equation}
\begin{aligned}
F_i(\mathbf{x})=-\Delta_{i}^{(+)} \phi(\mathbf{x}) + \epsilon_{ijk} \Delta_j^{(-)} L_k(\mathbf{x}) 
\end{aligned}
\end{equation}
where the scalar field $\phi$ has the same form as in eq.~(\ref{scalarfieldpbc}), with the Green's function replaced by the one for open boundary condition. The plaquette field $L$ also has the same form as in eq.~(\ref{Lpotentialpbc}), but the sum goes only over all plaquettes (not all lattice sites) and the Green's function $G_{\mathrm{plaq}}(\mathbf{x},\mathbf{y})$ is determined by a modified Laplace operator $\nabla^2_{\mathrm{plaq}}$ on the plaquettes: 

\begin{equation}
-\nabla^2_{\mathrm{plaq}} f(\mathbf{x})= 2d f(\mathbf{x}) - \sum_{i=1}^{d} (f(\mathbf{x}+\mathbf{\hat{e}}_i) + f(\mathbf{x}-\mathbf{\hat{e}}_i))
\end{equation}
where the difference is the constant factor  of $2d$, also at the boundaries, e.g. at $\mathbf{x}=0$: 

\begin{equation}
-\tilde{\nabla}^2 f(\mathbf{0}) = 2d f(\mathbf{0}) - \sum_{i=1}^d f(\mathbf{0}+\mathbf{\hat{e}}_i)
\end{equation}
All the above discussion applies immediately to the $d=2$ case, by embedding it in $d=3$.

\subsection{The shifts $s_{\mathbf{x},i}(\mathbf{y})$}

As a result of the Helmholtz decomposition, we obtain the shifts $s_{\mathbf{x},i}(\mathbf{y})$ discussed in section~\ref{dualformulation}, which describe the shifts in the electric plaquette variables $L(\mathbf{x})$ when a matter degree of freedom hops to an adjacent site. We just need to replace the general field $F_i(\mathbf{x})$ with a field which is zero everywhere and one on the link where hopping occurs. The resulting values for $L$ computed by eq.~(\ref{Lpotentialpbc}), (\ref{L2}) and (\ref{L3}) adapted to two dimensions give exactly the shifts $s_{\mathbf{x},i}(\mathbf{y})$ (analogously for open boundary conditions), e.g. for a shift in $\mathbf{\hat{e}}_1$-direction:

\begin{equation}
\begin{aligned}
s_{\mathrm{plaq},\mathbf{x},1}\left(\mathbf{y}\right) &= G\left(\mathbf{y},\mathbf{x}\right) - G\left(\mathbf{y},\mathbf{x}-\mathbf{\hat{e}}_2\right) \\ 
s_{\mathrm{const},\mathbf{x},1}\left(\mathbf{y}\right) &= \frac{1}{N^2} y_2 \\
s_{\mathbf{x},1}(1)=\frac{1}{N}
\end{aligned} 
\end{equation}
so that $s_{\mathbf{x},1}\left(\mathbf{y}\right)=s_{\mathrm{plaq},\mathbf{x},1}\left(\mathbf{y}\right)+s_{\mathrm{const},\mathbf{x},1}\left(\mathbf{y}\right)$ and with $s_{\mathbf{x},1}(1)$ the shift in the global loop variable $L_1$.


\section{Canonical commutation relations} \label{cancomrel}

In this section we show that the dual $B$/$L$-variables fulfill canonical commutation relations as stated in eq.~(\ref{comrelBL}). Using the expression of $L(\mathbf{y})$ in terms of the original electric field $E_i(\mathbf{y})$ on the links (see eq.~(\ref{LE})), the expression of $B(\mathbf{x})$ as the lattice curl of the gauge field $\phi_{j}(\mathbf{x})$ (see eq.~(\ref{Bdef})) and the original canonical commutation relations in eq.~(\ref{phiE}), we obtain

\begin{equation}
\begin{aligned}
\left[B(\mathbf{x}), L(\mathbf{y}) \right]&= \epsilon_{i j} \epsilon_{kl} \sum_{\mathbf{y'}} \Delta_{k,\mathbf{y}}^{(+)} G(\mathbf{y},\mathbf{y'})  
\left[ \Delta_{i,\mathbf{x}}^{(+)} \phi_{j}(\mathbf{x}),  E_{l}(\mathbf{y'}) \right] \\
&=\epsilon_{i j} \epsilon_{kl} \sum_{\mathbf{y'}} \Delta_{k,\mathbf{y}}^{(+)} G(\mathbf{y},\mathbf{y'})  
i \delta_{j l} \left(\delta_{\mathbf{x}+\hat{\mathbf{e}}_{i},\mathbf{y'}} - \delta_{\mathbf{x},\mathbf{y'}} \right) \\
&=\epsilon_{i j} \epsilon_{k j} i \Delta_{k,\mathbf{y}}^{(+)} \Delta_{i,\mathbf{x}}^{(+)} G(\mathbf{y},\mathbf{x})  \\
&=-i \Delta_{i,\mathbf{x}}^{(-)} \Delta_{i,\mathbf{x}}^{(+)} G(\mathbf{y},\mathbf{x}) \\
&=i  \delta_{\mathbf{x},\mathbf{y}}
\end{aligned}
\end{equation}
canonical commutation relations also for $B$ and $L$. In a completely analogous way one can derive the canonical commutation relations for $\theta$ and $M$.

\section{The modified Coulomb potential between the dual $M$-variables for periodic boundary conditions} \label{details}

If we consider periodic boundary conditions in the dual formulation in terms of the $\theta/M$-variables, the electric Hamiltonian gives rise to Coulomb-type interactions between the $M$-variables. The interaction potential $\widetilde{G}(\mathbf{x},\mathbf{y})$ is slightly modified compared to the potential for the matter degrees of freedom due to the global loops as discussed in section~\ref{dualformulation} in eq.~(\ref{extraCoulomb}). They change the Laplace operator (here in two dimensions) on the plaquettes to, see eq.~(\ref{MintermsofL}): 

\begin{equation}
\begin{aligned}
M(\mathbf{x})=&-\nabla^2 L (\mathbf{x}) + \epsilon_{ij} \Delta_{i}^{(+)} \delta_{x_i,0} L_j \\
&-\nabla^2 L (\mathbf{x}) + \left(\delta_{x_2,0} - \delta_{x_2,-1} \right) L_1 - \left(\delta_{x_1,0} - \delta_{x_1,-1} \right) L_2
\end{aligned}
\label{MLrealtion1}
\end{equation}
The plaquettes where the Laplace operator is altered are the ones sharing a link with one of the two axes. The relation between the global $M$-variables $M_1$ and $M_2$ and $L$ is 

\begin{equation}
\begin{aligned}
&M_1= N L_1+ \sum_{\substack{x_2=0 \\ x_1}} \left( L(\mathbf{x}) - L(\mathbf{x}-\mathbf{\hat{e}}_2) \right) \\
&M_2= N L_2- \sum_{\substack{x_1=0 \\ x_2}} \left( L(\mathbf{x}) - L(\mathbf{x}-\mathbf{\hat{e}}_1) \right)
\end{aligned}
\label{MLrelation2}
\end{equation}
If one defines an $M$-vector out of the plaquette variables $M(\mathbf{x})$ and the global variables $M_1$ and $M_2$, $M \equiv \left( M(\mathbf{x}), M_1,M_2 \right)$, and analogously for $L$, one can construct a system of linear equations for $M$ and $L$ out of eq.~(\ref{MLrealtion1}) and eq.~(\ref{MLrelation2}), denoted by $D$, i.e $D L \equiv M$, which can be inverted (after fixing some condition), resulting in  

\begin{equation}
L(\mathbf{x})= \sum_{\mathbf{y}} D^{-1}\left(\mathbf{x},\mathbf{y}\right) M(\mathbf{y}).
\end{equation}
Note that the sum over $\mathbf{y}$ also contains $1$ and $2$, corresponding to the global loop variables $M_1$ and $M_2$. Inserting this relation in the electric Hamiltonian in terms of $L$ in eq.~(\ref{etransL}), gives eq.~(\ref{extraCoulomb})
\begin{equation}
H_E^T=\frac{g^2}{2}\sum_{\mathbf{x},\mathbf{y}}M(\mathbf{x}) \hspace{1pt} \widetilde{G} \left(\mathbf{x},\mathbf{y}\right) M(\mathbf{y})
\end{equation}
with 

\begin{equation}
\begin{aligned}
\widetilde{G}(\mathbf{x},\mathbf{y})=\sum_{\mathbf{x'},i} &\left( \epsilon_{ij} \Delta^{(-)}_{i,\mathbf{x'}} D^{-1}\left(\mathbf{x'},\mathbf{x}\right) + \delta_{x_j,0} D^{-1}\left(i,\mathbf{x}\right) \right) \\ \times&\left( \epsilon_{ik} \Delta^{(-)}_{i,\mathbf{x'}} D^{-1}\left(\mathbf{x'},\mathbf{y}\right) + \delta_{x_k,0} D^{-1}\left(i,\mathbf{y}\right) \right) 
\end{aligned}
\end{equation}
with $i \neq j$ and $i \neq k$. Since $\mathbf{x}$ and $\mathbf{y}$ also include the global variables $M_1$ and $M_2$ (denoted by $1$ and $2$, as for example in $D^{-1}(i,\mathbf{x})$), there are non-trivial interactions between the global variables and the plaquette variables. For open boundary conditions, the above equation expression for $\widetilde{G}(\mathbf{x},\mathbf{y})$ reduces to $G(\mathbf{x},\mathbf{y})$.
\bibliography{ref}
\end{document}